\documentclass[pdflatex,sn-mathphys-num, ]{sn-jnl}% Math and Physical Sciences Numbered Reference Style
\usepackage{graphicx}%
\usepackage{multirow}%
\usepackage{amsmath,amssymb,amsfonts}%
\usepackage{amsthm}%
\usepackage{mathrsfs}%
\usepackage[title]{appendix}%
\usepackage{xcolor}%
\usepackage{textcomp}%
\usepackage{manyfoot}%
\usepackage{booktabs}%
\usepackage{algorithm}%
\usepackage{algorithmicx}%
\usepackage{algpseudocode}%
\usepackage{listings}%
\usepackage{bm}%

\newcommand{\overlap}[1]{\chi_{\mu}^{\dagger #1}\chi_{\nu}^{#1}} %writes overlap for derivations of rho, both large and small
\newcommand{\overlapsig}[1]{\chi_{\mu}^{\dagger #1}\sigma_c\chi_{\nu}^{#1}} %writes overlap for derivations of m_c, both large and small
\newcommand{\funcres}[1]{\int \frac{\delta \epsilon_{xc}}{\delta #1}} %result of functional, \rho_+/- , \gamma_++/--/+-

\newcommand{\matel}{D_{\mu\nu}^{TT}}

\newcommand{\dr}{d\mathbf{r} \ } %dr for integrals

\newcommand{\mathtxt}[1]{$\mathrm{#1}$}

\usepackage{xr}
\begin{document}

\title{A unified formalism for collinear and non-collinear approaches in the four-component Dirac-Kohn-Sham theory based on G-spinors}
%%=============================================================%%
%% GivenName    -> \fnm{Joergen W.}
%% Particle     -> \spfx{van der} -> surname prefix
%% FamilyName   -> \sur{Ploeg}
%% Suffix       -> \sfx{IV}

\author*[1,4]{\fnm{Giulia}\sur{Gamboni}}\email{giulia.gamboni@phd.units.it}

\author[2]{\fnm{Loriano} \sur{Storchi}}\email{loriano@storchi.org}

\author*[1,3]{\fnm{Paola} \sur{Belanzoni}}\email{paola.belanzoni@unipg.it}

\author*[3]{\fnm{Leonardo} \sur{Belpassi}}\email{leonardo.belpassi@cnr.it}

\affil*[1]{\orgdiv{Department of Chemistry, Biology and Biotechnology}, \orgname{University of Perugia}, \orgaddress{\street{via Elce di Sotto, 8}, \city{Perugia}, \postcode{06123}, \country{Italy}}}

\affil[2]{\orgdiv{Department of Pharmacy}, \orgname{University of Chieti-Pescara}, \orgaddress{\street{Via dei Vestini, 31}, \city{Chieti}, \postcode{ 66013}, \country{Italy}}}

\affil[3]{\orgdiv{Institute of Science and Chemical Technologies "Giulio Natta" - SCITEC }, \orgname{National Research Council, CNR}, \orgaddress{\street{via Elce di Sotto, 8}, \city{Perugia}, \postcode{06123},  \country{Italy}}}

\affil[4]{\orgdiv{Present address: Department of Chemical and Pharmaceutical Sciences}, 
\orgname{University of Trieste}, 
\orgaddress{\street{Via L. Giorgieri 1}, \city{Trieste}, \postcode{34127}, \country{Italy}}}


\abstract{Non-collinear density functional theories were developed to extend the use of established collinear exchange-correlation functionals to systems with unpaired electrons in the presence of significant spin-orbit coupling. These schemes were originally introduced mostly within relativistic two-component frameworks; a notable exception is the Dirac-Kohn-Sham (DKS) implementation in the ReSpect and BDF codes, which applies the non-collinear Scalmani-Frisch approach and non-collinear Generalized Gradient Approximation, respectively, at a fully four-component level.
A comparison of different approaches and implementations is not straightforward, as the methods are often formulated using different fundamental variables and numerical approximations. A consistent review of the formal and numerical aspects of collinear and non-collinear schemes has recently been reported (Desmarais et al., J. Chem. Phys. 154, 204110 (2021)) in the context of two-component methods.
In this work, we present an initial effort towards a unified formulation of collinear and non-collinear approximations, encompassing both canonical and Scalmani-Frisch schemes, within the relativistic four-component DKS formalism based on G-spinor basis sets. Our preliminary implementation of the collinear and canonical non-collinear formulations in the DKS module of the \texttt{BERTHA} code extends its applicability and provides a benchmark for a series of simple open-shell hydride molecules 
(namely, H$_2$X$^+$, with X = O, S, Se, Te, and Po).
Furthermore, we have analysed the well-known limitations of the collinear approach in the presence of spin-orbit coupling using some graphical tools to visualise the behaviour of the magnetisation vector derived from the four-component spinors. The analysis  provides a clear, qualitative explanation of the origin of the lack of energy rotational invariance and its enhancement on descending along the periodic table. 
Finally, we show that incorporating the magnetisation vector into the reformulated non-collinear canonical LDA approach enables a description of H$_2$ dissociation - and open-shell systems more broadly - that closely parallels unrestricted non-relativistic approaches, notably without explicitly imposing the broken symmetry solution as is often required in non-relativistic collinear calculations. This unified formulation forms the basis for a rigorous comparison between different numerical approximations, which will be essential for obtaining stable results for the non-collinear GGA exchange-correlation functionals.}

%\abstract{This work presents a unified formulation of collinear and non-collinear (canonical and Scalmani-Frisch) approximations within a fully relativistic, four-component Dirac-Kohn-Sham (DKS) framework using G-spinor basis sets. Implemented in the BERTHA code, the method is benchmarked on open-shell hydride molecules (H$_2$X$^+$, with X = O, S, Se, Te, and Po). By graphically visualizing the spinor-derived magnetization vector, we provide a clear physical explanation for the lack of energy rotational invariance inherent in collinear approaches under strong spin-orbit coupling. Furthermore, our reformulated non-collinear canonical LDA accurately describes H$_2$ dissociation and open-shell systems without requiring explicitly imposed broken-symmetry solutions. This unified framework establishes a rigorous foundation for evaluating numerical approximations and developing stable, non-collinear Generalized Gradient Approximation (GGA) functionals.}


\maketitle

\section{Introduction}
Molecular systems with unpaired electrons and observable spin-orbit coupling (SOC) effect, due to the presence of heavy elements, play a crucial role in various fields. In the recent search for new quantum computing hardware \cite{hardware_book}, electron spins are used as the fundamental physical quantity to define qubits. Growing interest is being given to molecular qubits \cite{bayliss_1,laorenza_image,bayliss_2,Baldinelli2025}, due to their bottom-up synthesis and tunable characteristics.
Chirality Induced Spin Selectivity \cite{CISS_review} is a novel effect observed in the chiral surfaces and molecular systems, where an electronic current is polarized passing through the material, with applications ranging from spintronics \cite{Michaeli_2017} to enantioselectivity\cite{enantio_surf}. 
Finally, spin-forbidden reactions \cite{harvey2001spin,B200675H,mecp_ts,ricciarelli,ricciarelli_2,sorbelli} are among the most notable examples where the inclusion of spin-orbit coupling (SOC) is essential for accurately describing the potential energy surface and activation parameters of the reactions.
Beyond ground-state properties, the treatment of electronically excited states in the presence of spin-orbit coupling represents a major frontier in quantum chemistry. In the context of Density Functional Theory (DFT), by extending the four-component Dirac-Kohn-Sham (DKS) equations to the time-dependent domain (TD-DKS), it becomes possible to directly probe spin-forbidden transitions, core-level excitations, and intense spin-orbit coupling effects in heavy-element complexes without relying on perturbative treatments. While the present work focuses on the foundational ground-state collinear and non-collinear formalisms, their robust implementation is a prerequisite for accurate TD-DFT treatments of the electronic excited states.

When using a non-relativistic, spin-independent Hamiltonian, spin is a good quantum number, meaning that in the treatment of molecular systems with unpaired electrons one can adopt a separate description of spin-up and spin-down electrons. This leads to the so-called spin-unrestricted formulations of the KS-DFT method. For a detailed analysis of these two approaches, we refer the reader to Ref. \cite{spinDFT}. In this non-relativistic context, the exchange-correlation (\textit{xc}) energy is a functional of the total electron density and of one Cartesian component of the magnetisation that can be freely chosen. The approach, referred to as collinear, results insufficient for describing the system if the Hamiltonian is spin-dependent, for instance, if it includes a spin-orbit-coupling (SOC) term. Perhaps the most evident problem is that the energy of a molecular system with an unpaired electron is not invariant with respect to a rotation in space. In the presence of spin-dependent Hamiltonian, the rotational transformation can be considered as a rotation in "real" space, followed by a rotation in "spin" space. The first does not affect the energy of the system, but the lack of commutation between the spin and the Hamiltonian results in an energy variation after the "spin" rotation. This break in rotational invariance of the collinear approach is of course highly undesirable when describing molecular systems \cite{van_wullen2002}.

Another limitation of the method is its inability to study spin dynamics: by arbitrarily fixing the quantization axis of the spins they cannot evolve properly over time, making it impossible to study spin-relaxation phenomena accurately. In a real-time calculation, no variation in their direction would be observed, as they are fixed \textit{a priori} along one axis. For example, it is well known that the single-triplet transition energy is significantly overestimated \cite{DeSantis2020}. Clearly, new exchange-correlation density functionals designed to account for noncollinearity are highly desirable, and significant effort is underway in this direction.
A complication in the derivation of a suitable exchange-correlation functional arises from the fact that the fundamental variables of the theories depend on the level at which relativistic effects are included. For example, in a fully relativistic four-component framework, the appropriate formulation is four-current DFT, and the fundamental variable of the theory is the covariant four-current density \cite{ENGEL2002523}. In a two-component formulation, the relevant theory is spin-current DFT, which formally depends on a total of sixteen variables: the electron density, magnetisation, orbital-current density, and nine spin-current densities. An active research activity is underway to extract exact conditions of non-collinear functionals from  internal symmetries invariance considerations (see, for instance, Ref.\cite{Desmarais:2026}).

A common strategy to generalise DFT to the non-collinear scheme has been to pragmatically use standard non-relativistic (spin-polarised) functionals. Originally introduced by Kübler et al.\cite{JKubler_1988} for the local density approximation (LDA) functional in the context of two-component generalised DFT, the non-collinear scheme is based on the idea that the quantisation axis of the spin is no longer fixed in space but instead varies locally, adopting the direction of the magnetisation density.
In this approach, known as the canonical scheme, the spin-polarised densities are redefined via the modulus of the magnetisation, rather than with respect to a chosen component as in the collinear scheme. Thus, even if the \textit{xc} functional maintains the same form, due to the redefinition of the spin-polarised densities, it now depends on the density and the modulus of the magnetisation ($m$).
Its extension to Generalised Gradient Approximation (GGA) or meta-GGA functionals shows significant problems of numerical stability: in current implementations one typically removes numerically unstable terms and adopts specific tricks that introduce difficulties in the systematic reproduction of data between different software.
A slightly different strategy underlies the scheme proposed by the Scalmani-Frisch approach \cite{Scalmani2012}, which represents a step forward in the search for more numerically stable non-collinear methods. It differs from the canonical only in the definition of the $\gamma_{\pm \pm}$ variables in the GGA exchange-correlation functional, the two methods are otherwise identical in the LDA formulation.
The Scalmani-Frisch is now the most widely implemented method for handling non-collinear spins \cite{Egidi2017}.

Recently, an additional scheme called multicollinear \cite{multicollinear} has been presented, in which the local magnetisation vector is partitioned into domain-specific axes rather than relying on a single global or strictly local quantisation axis. This approach aims to restore rotational invariance while mitigating the numerical instabilities typically encountered in standard non-collinear formulations. It appears to be very promising; however, we do not treat it explicitly in the present work.
When heavy elements are involved, the accurate inclusion of SOC effects is essential: the Dirac-Kohn-Sham (DKS) theory \cite{PhysRevB.7.1912,AKRajagopal_1978} offers an accurate method, using the four-component formalism of the Dirac equation within the DFT framework to include all exchange-correlation many-body effects. However, treatment of unpaired spins within the DKS theory is not straightforward. Unfortunately, standard \textit{xc} functionals remain predominantly non-relativistic in design, requiring the mapping of relativistic densities onto non-relativistic functionals defined in terms of polarised densities (i.e. of \mathtxt{\alpha} and \mathtxt{\beta} electrons). 
%Current methodologies encompass collinear \cite{van_wullen2002}, non-collinear (canonical %\cite{van_wullen2002} and Scalmani-Frisch \cite{Scalmani2012,Egidi2017} ) or %multicollinear approximations, depending on how the magnetisation vector, derived from %the Gordon decomposition of the four-current \cite{spinDFT}, enters in the redefinition %of the functional variables. 

We mention that the first formulation and numerical applications of the collinear scheme  and canonical non-collinear in the context of four-component relativistic DFT is due to Anton et al.\cite{Anton:2002,Engel2004}. The implementation was carried out in a molecular code which used a multipole multicenter expansion scheme to evaluate the electronic Hartree potential and the numerical issues were not investigated. 
In 2003, Wang and Liu\cite{Wang2003_Polarization} presented a systematic comparison of different spin-polarisation schemes in open-shell atoms. Significant effort has been made to include non-collinear kernels in the linear response using the Dirac-Coulomb Hamiltonian to describe, for instance, accurate transitions in systems containing heavy elements (see Refs.\cite{https://doi.org/10.1002/qua.22065, Li01122013}).
Of particular relevance to this article is the work by Komorovsky, Cherry and Repisky, who present the formulation of ground state and linear response four-component Dirac-Coulomb Hamiltonian based on the non-collinear scheme developed by Scalmani and Frisch, implemented in the ReSpect code \cite{10.1063/1.5121713}. As far as we know, there is no evidence in the literature that collinear and non-collinear methods have been systematically investigated in molecules at the fully relativistic four-component level.

The scope of this work is to reformulate the collinear and non-collinear (canonical and Scalmani-Frisch) approaches within four-component (DKS) theory based on G-spinors, as implemented in the BERTHA programme. In particular, we derive the basic expressions for their matrix elements in G-spinor basis set. The exact constraints of the exchange-correlation functionals are discussed for both non-collinear canonical and Scalmani-Frisch approaches in relation to the local and global torque theorems.

We also extend the applicability of the DKS module of BERTHA, which until now was based on the "density-only" approximation, by implementing collinear (LDA and GGA) and canonical non-collinear (LDA) functionals, and test the implementation on a series of chalcogen hydrides, $\mathrm{H_2X}$ ($\mathrm{X = O, S, Se, Te, Po}$), which exhibit a systematically increasing spin-orbit coupling (SOC) effect. Comparative calculations were also performed using the PySCF \cite{PYSCF} and ORCA \cite{ORCA5} software packages.
We further demonstrate with numerical examples that both the collinear and non-collinear methods, as expected, reduce to non-relativistic unrestricted-KS, thereby inheriting its characteristic spin contamination. As a preliminary implementation, this framework serves as a basis for several ongoing and future developments, including the implementation of GGAs alongside solutions to their associated numerical instabilities, the integration of the Scalmani-Frisch approach, and their incorporation into our real-time implementation \cite{DeSantis2020} to enable explicit spin dynamics calculations. The methodologies presented may use density fitting algorithms to reduce overall computational cost.

This work was carried out within the framework of the Theoretical Chemistry and Computational Modelling (TCCM) program. The content of this paper is organised as follows: Section 2 provides a brief overview of relativistic four-component DKS theory within the G-spinor basis of the BERTHA program, including the formal definitions of the spin density and magnetisation vector. It presents the explicit mathematical formulae for the collinear and non-collinear methods, focusing on the canonical and Scalmani-Frisch approaches and discusses the specific expressions for the underlying matrix elements. In Section 3, to demonstrate the efficacy of this implementation, we report results on the ionisation energies of heavy hydride molecules ($\text{H}_2\text{X}^+$, where $\text{X} = \text{O}$, $\text{S}$, $\text{Se}$, $\text{Te}$, and $\text{Po}$) using both collinear and canonical non-collinear LDA functionals, and analyse the instructive benchmark of the $\text{H}_2$ dissociation curve. Conclusions and perspectives are presented in Section 4.

\section{Theory and derivations}

Before describing the Dirac-Kohn-Sham (DKS) implementation in the BERTHA code using the G-spinor basis and outlining the formulae derived for the exchange-correlation matrix elements in both collinear and non-collinear forms, it is useful to briefly summarise the key aspects of the relativistic DKS method. The aim of the following overview is to highlight the specific approximations typically employed in modern relativistic computations.

In the original extension of the Hohenberg-Kohn (HK) theorem to relativistic systems proposed by Rajagopal et al. \cite{PhysRevB.7.1912,AKRajagopal_1978,MacDonald1979}, the ground-state four-current density $j^{\mu}(\mathbf{r})$ plays a role analogous to the ground-state electron density $\varrho(\mathbf{r})$ in the non-relativistic context. Specifically, $j^{\mu}(\mathbf{r}) = (c\varrho(\mathbf{r}), \mathbf{j}(\mathbf{r}))$ serves as the fundamental variable that uniquely determines the many-particle ground state up to a gauge-transform phase factor. Consequently, the ground-state total energy $E_{\text{tot}}$ can be expressed as a unique functional of $j^{\mu}$, implying the existence of a local, effective exchange-correlation four-potential that accounts for all many-body interactions.

Engel and Dreizler \cite{engel2011density} emphasised that a rigorous proof within quantum electrodynamics (QED) must be formulated in terms of the renormalised ground-state energy and four-current, ensuring that the renormalisation scheme remains compatible with the HK existence theorem for relativistic DFT (RDFT). Addressing renormalisation is equally critical when deriving approximations for the exchange-correlation functional $E_{\text{xc}}[j]$ \cite{engel2011density}, an intricate task that has not yet been addressed in practical quantum chemistry implementations. Instead, electronic structure calculations typically employ the no-pair approximation. Within this framework, all effects arising from the creation of virtual electron-positron pairs are neglected, yielding a theory in which both the total charge and the particle number are conserved.
The ground state four-current density $j^{\mu}({\bf r})=(c\rho(\bf r),{\bf j(\bf r)})$ of a relativistic Kohn-Sham system is defined as
the sum of single particle four-spinors ($\Psi_i({\bf r})$) associated with the positive-energy states (electrons).
These spinors take the form
\begin{equation}
{\bf \Psi}_i({\bf r})= \left[ \begin{array}{c}
                          \Psi_i^{(1)}({\bf r}) \\
                          \Psi_i^{(2)}({\bf r}) \\
                          \Psi_i^{(3)}({\bf r}) \\
                          \Psi_i^{(4)}({\bf r}) 
                        \end{array} \right],
\end{equation}
where the charge density ($\varrho(\bf r)$) is  evaluated as the scalar product of four-component spinors according to:
\begin{equation}\label{density}
\varrho({\bf r})=\sum_a {\bf \Psi}_{a}{\bf (r)}^{\dagger}{\bf \Psi}_{a}{\bf (r)}=\sum_a \varrho_a({\bf r})
\end{equation} while the spatial current density,
${\bf j}=(j_x,j_y,j_z)$,
can be expressed  as:
\begin{equation}
{\bf j} = c\sum_a {\bm \Psi}_a^\dagger({\bf r})
\bm{\alpha}{\bm \Psi}_a({\bf r})
\end{equation}
where $c\bm{\alpha}=(c \alpha_x,c\alpha_y,c\alpha_z)$
denotes, as above, the $4\times 4$ matrix representation of the
relativistic electron current operator.
These single particle spinors are solutions of the relativistic KS equation, which in its general form reads:
\begin{equation}
\{c \mbox{\boldmath$\alpha$\unboldmath} \cdot {\bf p}+ {\bf \beta} c^2+v^L({\bf r}) + \mbox{\boldmath$\alpha$\unboldmath} \cdot {{\bf v}^T} ({\bf r}) \}{\bf \Psi}_{i}{\bf (r)}=E_i{\bf \Psi}_{i}{\bf (r)}.
\end{equation}
Here $c$ is the speed of light in vacuum, $\bf p$ is the electron
momentum and $\bm \alpha$ and $\beta$ are Dirac matrices defined as:
\begin{equation}
{\bm \alpha} = \left(
\begin{array}{cc}
0 & {\bm \sigma} \\
{\bm \sigma} & 0
\end{array}
\right)\ \mbox{and}\ 
{\bm\beta} =
\left(
\begin{array}{cc}
\mathbf{I} & 0 \\
0 & -\mathbf{I}
\end{array}
\right)
\end{equation}
where ${\bm\sigma}=(\sigma_x,\sigma_y,\sigma_z)$, $\sigma_q$ are the
$2\times 2$ Pauli spin matrices and $\mathbf{I}$
is a $2\times 2$ identity matrix. The diagonal longitudinal potential operator $\upsilon^L({\bf r})$ is given by the sum of three terms:
\begin{equation}\label{eq:vlpot}
\upsilon^L({\bf r})=
\upsilon_{ext}({\bf r})+\upsilon^L_{H}({\bf r})+\upsilon^L_{xc}({\bf r}).
\end{equation}
Here, $\upsilon_{ext}({\bf r})$ represents the external potential
due to the fixed nuclei, and
\begin{equation}
\upsilon^L_{H}[\varrho({\bf r})]=\int\,d^3s\frac{\varrho({\bf s})}{|{\bf r}-{\bf s}|},
\end{equation}
is the electronic Coulomb interaction, which is a functional of
the relativistic charge density, $\varrho({\bf r})$.

The transverse potential operator ${\bf v}^T({\bf r})$ is given by
\begin{equation}
{\bf v}^T({\bf r})=
{\bf A}_{ext}({\bf r})+{\bf v}^T_{H}({\bf r})+{\bf v}^T_{xc}({\bf r}).
\end{equation}
${\bf A}_{ext}({\bf r})$ represents the external vector field 
due to a magnetic field and 
\begin{equation}
{\bf v}^T_{H}(\bf r)=\int\,d^3s\frac{\bf j({\bf s})}{|{\bf r}-{\bf s}|},
\end{equation}
represents the magnetic electronic interaction (which incorporates the Breit interaction) as a functional of the spatial  current density, $\bf j({\bf r})$. 
The terms $\upsilon^L_{xc}({\bf r})$ and ${\bf v}^T_{xc}({\bf r})$ are defined as functional derivatives of the exchange-correlation functional, $E_{xc}[j]$, with respect to the electron density and the spatial component of the four-current density, respectively:
 \begin{equation}
\upsilon^L_{xc}({\bf r})=\frac{\delta E_{xc}[j]}{\delta \rho}; \hspace{1cm} 
{\bf v}^T_{xc}({\bf r})=\frac{\delta E_{xc}[j]}{\delta \bf j}.
\end{equation}

These functionals implicitly include the full magnetic interaction as well as electron correlation effects. However, these formal expressions are rarely useful in practice, because the exact \textit{xc} functionals are unknown and general, accurate approximations do not yet exist. A significant simplification arises in the absence of an external magnetic field, when the nuclei are fixed and nuclear magnetic moments are neglected ($\mathbf{A}_{\text{ext}}(\mathbf{r}) = 0$). In this case, the external four-potential reduces to the scalar nuclear potential, $v_{\text{ext}}(\mathbf{r})$. Analogously to the non-relativistic KS scheme, a one-to-one mapping exists between $v_{\text{ext}}(\mathbf{r})$, the ground-state wavefunction, the total energy, and the electron density $\varrho(\mathbf{r})$. The spatial current density can therefore be treated as an implicit functional of the density ($\mathbf{j}[\varrho]$), allowing the total energy to be written as $E[j[\varrho]]$. Consequently, the four-component potential in the DKS equations simplifies to a local scalar potential that depends on $\delta E_{\text{xc}}[\varrho, \mathbf{j}]/\delta \varrho$ and the contraction of $\delta E^{\text{T}}_{\text{H}}[j]/\delta \mathbf{j}$ and $\delta \mathbf{j}[\varrho]/\delta \varrho$ (see for instance Eq. 25 in Ref.\cite{Engel-hand-book:2015}).
Since the exact functional dependence $\delta \mathbf{j}[\varrho]/\delta \varrho$ is unknown, it is standard practice to neglect the explicit $\mathbf{j}$-dependence of both $E^{\text{T}}_{\text{H}}$ and $E_{\text{xc}}$. This is known as the "density-only" approximation of the DKS scheme. This approximation is reasonably accurate for closed-shell systems in the absence of external magnetic fields, where time-reversal symmetry is preserved and the spatial current vanishes ($\mathbf{j} = \mathbf{0}$). Under these conditions, the exchange-correlation functional depends only on the total density, allowing standard non-relativistic functionals to be reasonably applied.
Conversely, for open-shell or spin-polarized molecular systems, the density-only approximation is highly unsatisfactory. To bridge the gap between the full four-current approach and the limiting density-only DKS scheme, one can devise relativistic spin-DFT intermediate formulations.
If one assumes that the system couples with an external magnetic field ($\bf B_{ext}$) exclusively from the magnetisation density (rather than $\bf j \cdot {\bf A}_{ext}({\bf r})$), then a one-to-one mapping can be established between the external magnetic field and the magnetisation vector ($\bf m$). Relativistic "spin" DFT is therefore an approximation to the full four-current approach that neglects the coupling between the orbital part of the current and the external field. When the external magnetic field is zero, this formulation is formally exact: the total energy and all observables, including the spatial current, can be expressed as functionals of $\rho$ and the magnetisation vector (e.g. $E[\rho, \bf m], \bf j[\rho, \bf m]$). Assuming a noninteracting system reproduces the ground state $\rho$ and $\bf m$ of the interacting system, one can devise a relativistic spin DKS equation, where the magnetisation of the non-interacting system is given by 
\begin{equation}
    \mathbf{m}(\mathbf{r}) = \sum_i \Psi^\dagger_i(\mathbf{r})\mathbf{\Sigma}\Psi_i(\mathbf{r}),
\end{equation}
where 
$\mathbf{\Sigma}=(\Sigma_x,\Sigma_y,\Sigma_z)$ with
\begin{equation}
{\bm \Sigma} = \left(
\begin{array}{cc}
{\bm \sigma} & 0 \\
0 & {\bm \sigma} 
\end{array}
\right)\
\end{equation}
where ${\bm\sigma}=(\sigma_x,\sigma_y,\sigma_z)$ are
the $2\times 2$ spin matrices Pauli mentioned earlier.
The definition arises by the Gordon decomposition \cite{spinDFT}, in which the spatial components of four-current density are decomposed into an orbital current and the curl of $\bf m$ ($\nabla \times \bf m$), the latter retains the explicit spin-dependent part. 
%Note that in the Gordon decomposition the $\bf \Sigma$ matrices are multiplied by the diagonal matrix $\beta$; both conventions appear in the literature. 
While some authors include the diagonal Dirac $\mathbf{\beta}$ matrix inside the definition of the magnetisation operator \cite{gohr_phdthesis, spinDFT}, it is omitted in this formulation. The numerical differences resulting from this choice have been demonstrated to be negligible for chemical systems \cite{https://doi.org/10.1002/qua.22065}.

The relativistic spin DKS equations reads as:
\begin{equation}
\{c \mbox{\boldmath$\alpha$\unboldmath} \cdot {\bf p}+ {\bf \beta} c^2+\upsilon^s({\bf r}) + \mbox{\boldmath$\Sigma$\unboldmath} \cdot \bf B_s\}{\bf \Psi}_{i}{\bf (r)}=\epsilon_i{\bf \Psi}_{i}{\bf (r)},
\end{equation}
where 
\begin{equation}
\upsilon^s({\bf r})=\upsilon_{ext}({\bf r})+\upsilon^L_{H}({\bf r})+
\frac{\delta E_H^T[\bf j[\rho, \bf m]]}{\delta \rho(r)}
+\frac{\delta E_{xc}[\rho, \bf m]}{\delta \rho(r)}
\end{equation}
and 
\begin{equation}
{\bf B_s(r)}={\bf B_{ext}}({\bf r})+
\frac{\delta E_H^T[\bf j[\rho, {\bf m}]]}{\delta \bf m(r)}
+\frac{\delta E_{xc}[\rho, \bf m]}{\delta \bf m(r)}.
\end{equation}
In practical implementations, the transverse contribution to the Hartree potential is typically neglected. The central quantity of the theory becomes the exchange-correlation functional, $E_{\text{xc}}[\varrho, \mathbf{m}]$, which generally depends on all the three components of the magnetisation vector field. Explicit forms for these three-component vector functionals are generally not available.

If one assumes that the coupling between the external magnetic field and the system can occur only via a single component of the magnetisation, then the energy becomes a functional of $\rho$ and this component of the magnetisation vector ($E_{tot}[\rho, m_z]$ and $E_{xc}[\rho, m_z]$). This defines the collinear approach, whose non-relativistic limit corresponds to standard spin-density functional theory. 
%The functional can also be expressed in terms of spin-density ($\rho_\pm = \frac{1}{2}(\rho \pm m_z)$). This formulation is highly advantageous because approximate exchange-correlation functionals developed in the non-relativistic domain can be directly used in practical applications.

Alternatively, if one requires the non-interacting system to reproduce the local vector magnitude of the magnetisation field $|\mathbf{m}(\mathbf{r})|$ at each point in space, rather than a fixed component $m_z(\mathbf{r})$, one arrives at the non-collinear approach. In this case, the magnetisation forms a vector field whose orientation varies continuously with spatial position. %Here, the functional is defined in terms of generalised spin densities as:
%\begin{equation}
%\varrho_\pm = \frac{1}{2}\left(\varrho \pm |\mathbf{m}|\right).
%\end{equation}
In the simple case where no external magnetic field is present and the transverse Hartree interaction $E_{\text{H}}^{\text{T}}$ is neglected, the local potential term $\mathbf{\Sigma} \cdot \mathbf{B}_{\text{s}}$ simplifies to $\Sigma_z \frac{\delta E[\varrho, m_z]}{\delta m_z}$ for the collinear model, whereas for the non-collinear model it is given by:
\begin{equation}\label{eq:noncollinear}
\mathbf{\Sigma} \cdot \mathbf{B}_{\text{s}} = \sum_{c=x,y,z} \Sigma_c \frac{\delta E[\varrho, |\mathbf{m}|]}{\delta m_c},
\end{equation}
which involves all three Cartesian components of the magnetisation.
More details on the collinear and non-collinear methods are given in the sections below.

\subsection{Four-Component Formalism of the Dirac-Kohn-Sham Equations Using a G-spinor Basis Set}

The primary objective of this work is to derive explicit formulations for collinear and non-collinear exchange-correlation matrix elements suitable for direct implementation within the BERTHA program framework. Before detailing these new derivations, it is useful to summarise the current DKS approximations implemented in the BERTHA code. For extensive discussions on recent computational advancements in the DKS module of BERTHA, such as density fitting schemes, parallelisation strategies, and GPU acceleration, the reader is referred to specific reviews \cite{C1CP20569B,10.1063/5.0002831, StorchiBelpassiGPU2025}.
Here, we highlight only the essential features relevant to the current discussion.
The standard version of the DKS equations implemented in the BERTHA code is based on the approximation 
that considers only the longitudinal electrostatic interactions. Following the discussion in Ref. \cite{QuineyBelanzoni2002}, the simplified DKS equation reads:
\begin{equation}\label{Dirac}
\{c \mbox{\boldmath$\alpha$\unboldmath} \cdot {\bf p}+ {\bf \beta} c^2+\upsilon^L({\bf r})\}{\bf \Psi}_{i}{\bf (r)}=E_i{\bf \Psi}_{i}{\bf (r)}.
\end{equation}

This formulation corresponds to the "density-only" approximation introduced above, where $v^{\text{L}}[\varrho(\mathbf{r})]$ is the longitudinal potential defined in Eq.~\eqref{eq:vlpot}, the Breit interaction is neglected, and $v^{\text{L}}_{\text{xc}}[\varrho(\mathbf{r})]$ represents the relativistic longitudinal exchange-correlation potential associated with the longitudinal exchange-correlation energy $E^{\text{L}}_{\text{xc}}[\varrho(\mathbf{r})]$.
Since the exact functional form of $E^{\text{L}}_{\text{xc}}[\varrho(\mathbf{r})]$ remains unknown, it must be approximated. 
In BERTHA, standard non-relativistic density functionals (such as the LDA or GGA) are employed as a foundational starting point, onto which relativistic corrections can be 
appended. 
The total electronic energy is then expressed as:
\begin{equation}\label{tote}E_{\text{tot}} = \sum_{i}\varepsilon_i - E_{\text{H}}^{\text{L}}[\varrho(\mathbf{r})] + E_{\text{xc}}^{\text{L}}[\varrho(\mathbf{r})] - \int v_{\text{xc}}^{\text{L}}[\varrho(\mathbf{r})] \varrho(\mathbf{r}) d\mathbf{r},
\end{equation}
where the summation extends over the occupied positive-energy bound states (electronic states), 
and $E_{\text{H}}^{\text{L}}[\varrho(\mathbf{r})]$ is the electronic Coulomb energy defined by:
\begin{equation}E_{\text{H}}^{\text{L}}[\varrho(\mathbf{r})] = \frac{1}{2}\int v_{\text{H}}^{\text{L}}[\varrho(\mathbf{r})]\varrho(\mathbf{r}) d\mathbf{r}.
\end{equation}
Within the BERTHA code, the one-particle spinors are expanded in a basis set of G-spinors \cite{C1CP20569B}.
These are two-component spin-orbit coupled objects derived from the spherical Gaussian-type function (SGTF) 
basis, structured as:
\begin{equation}
    \Psi_i = 
\left [
\begin{array}{c}
\sum_{\mu=1}^N c_{i\mu}^L \chi_\mu^{(L)}(\mathbf{r}) \\
i\sum_{\mu=1}^N c_{i\mu}^S \chi_\mu^{(S)}(\mathbf{r})
\end{array}
\right ]
\end{equation}
where $N$ is the number of basis functions, and $c_{i\mu}^{\text{T}}$ (with $\text{T} = \text{L}, \text{S}$ 
denoting the large and small components, respectively) represent the expansion coefficients.
G-spinors do not suffer from the variational collapse and satisfy the kinetic balance
constraint (see Ref.~\cite{Dyall1990} and references therein). 
Furthermore, for the evaluation of multicenter integrals, they preserve the key advantages that make Gaussian-type functions standard in non-relativistic quantum chemistry.
Designed to preserve  the transformation properties of central-field atomic four-spinors, 
G-spinors are simultaneous eigenfunctions of the total angular momentum operators $\hat{\mathbf{j}}^2$ and $\hat{j}_z$, 
as well as the Johnson-Lippmann fine-structure operator \cite{Quiney:2002}, with eigenvalues $\kappa=\pm 1, \pm 2, \pm 3, \dots$ (for an explicit algebraic definition, see Ref.~\cite{grant07}, p. 544).
In the G-spinor representation, the large- and small-component density matrices are defined as:
\begin{equation}\label{DTT}D^{(\text{TT}')}_{\mu \nu}=\sum_i c^{(\text{T})\ast}_{i\mu} c^{(\text{T}')}_{i\nu},
\end{equation}
where the index $i$ runs over all occupied positive-energy states. The total electron density is given by:
\begin{equation}
\label{eq:totden1}
\varrho(\mathbf{r}) = \sum_{\text{T}=\text{L},\text{S}}\sum_{\mu,\nu} D_{\mu\nu}^{(\text{TT})} \chi_\mu^{\text{T}\dagger}(\mathbf{r})\chi_\nu^{\text{T}}(\mathbf{r}),\end{equation}while the spatial components of the current density read:\begin{equation}j_c(\mathbf{r}) = \sum_{\text{T}\neq \text{T}'}\sum_{\mu,\nu} D_{\mu\nu}^{(\text{TT}')} j^{(\text{TT}')}_{c,\mu\nu}(\mathbf{r}) \quad (c=x,y,z),
\end{equation}
where $j^{(\text{TT}')}_{c,\mu\nu}(\mathbf{r})$ represents the G-spinor overlap current densities.
The matrix representation of the DKS operator is given by the block form:
\begin{equation} \label{DKSmatrix}
{\bf H}_{DKS}=
\left[ \begin{array}{cc}
{\bf V}^{(LL)} + mc^2 {\bf S}^{(LL)} &  c{\bm{\Pi}}^{(LS)} \\
c{\bm{\Pi}}^{(SL)}  & {\bf V}^{(SS)} - mc^2 {\bf S}^{(SS)}
\end{array}\right]
\end{equation}
where the diagonal blocks are constructed as $\mathbf{V}^{(\text{TT})} = \mathbf{v}^{(\text{TT})}+\mathbf{J}^{(\text{TT})}+\mathbf{K}^{(\text{TT})}$.
The corresponding matrix eigenvalue equation is:
\begin{equation}
{\bf H}_{DKS}
\left[ \begin{array}{c} {\bf c}^{(L)} \\ {\bf c}^{(S)}
\end{array}\right]
= E \left[ \begin{array}{cc}
{\bf S}^{(LL)} & 0  \\ 0  & {\bf S}^{(SS)}
\end{array}\right]
\left[ \begin{array}{c} {\bf c}^(L) \\ {\bf c}^(S) \end{array}\right]
\end{equation}
The  matrices $\mathbf{v}^{(\text{TT})}$, $\mathbf{J}^{(\text{TT})}$, $\mathbf{K}^{(\text{TT})}$, 
$\mathbf{S}^{(\text{TT})}$, and $\mathbf{\Pi}^{(\text{TT}')}$ represent the nuclear,
Hartree, exchange-correlation, overlap, and kinetic momentum operators, respectively. 
Their individual matrix elements are explicitly defined by:
\begin{eqnarray}
v_{\mu\nu}^{(TT)} &=&
\int v_{N}({\bf r}) \rho_{\mu\nu}^{(TT)}({\bf r})\,d{\bf r} \\
J_{\mu\nu}^{(TT)} &=&
\int v_{H}^{(l)}[\rho({\bf r})]\rho_{\mu\nu}^{(TT)}({\bf r})\,d{\bf r}
\label{jmatrix2bis} \\
K_{\mu\nu}^{(TT)} &=&
\int v_{xc}^{(l)}[\rho({\bf r})]\rho_{\mu\nu}^{(TT)}({\bf r})\,d{\bf r}
\label{kmatrix2bis} \\
S_{\mu\nu}^{(TT)} &=& \int \rho_{\mu\nu}^{(TT)}({\bf r})\,d{\bf r} \\
\Pi_{\mu\nu}^{(TT')} & = & \int
                  M_{\mu}^{(T)\dagger}({\bf r})
                  \left(\bm{\sigma} \cdot {\bf p}\right)
                  M_{\nu}^{(T')}({\bf r})\,d{\bf r}
\end{eqnarray}

Because $\mathbf{H}_{\text{DKS}}$ depends implicitly on the density matrices through $v_{\text{xc}}^{\text{L}}[\varrho]$ and $v_{\text{H}}^{\text{L}}[\varrho]$, these equations must be solved iteratively to self-consistency.
The Coulomb ($\mathbf{J}^{(\text{TT})}$) and exchange-correlation ($\mathbf{K}^{(\text{TT})}$) matrices 
contribute to the diagonal blocks. 
The calculation of $\mathbf{J}^{(\text{TT})}$ involves evaluating six-dimensional two-electron repulsion integrals, which are computed analytically using a relativistic generalization of the standard non-relativistic $J$-matrix algorithm \cite{Quiney:2002, C1CP20569B}. 
Conversely, the exchange-correlation elements $\mathbf{K}^{(\text{TT})}$ are evaluated via numerical integration. Multicenter integrals involving the total density $\varrho(\mathbf{r})$ and its gradient field $|\nabla\varrho(\mathbf{r})|$ are evaluated using Becke-style fuzzy  partitioning schemes \cite{2bisBecke:88}.
We note that non-local exact Fock exchange contributions (omitted in this work) would introduce non-vanishing off-diagonal blocks ($\text{T} \neq \text{T}'$). In the G-spinor representation, the fundamental physical observables are the electron density,
\begin{equation}
\varrho(\mathbf{r}) = \sum_{\text{T}=\text{L},\text{S}}\sum_{\mu\nu}D_{\mu\nu}^{\text{TT}}\chi_\mu^{\text{T}\dagger}(\mathbf{r})\chi_\nu^{\text{T}}(\mathbf{r}),
\end{equation}
and the magnetisation vector field,
\begin{equation}\mathbf{m}(\mathbf{r}) = \sum_{\text{T}=\text{L},\text{S}}\sum_{\mu\nu}D_{\mu\nu}^{\text{TT}}\chi_\mu^{\text{T}\dagger}(\mathbf{r}) \mbox{\boldmath$\sigma$\unboldmath}\chi_\nu^{\text{T}}(\mathbf{r}),\end{equation}
where $\mbox{\boldmath$\sigma$\unboldmath}$ represents the vector of $2\times2$ Pauli matrices ($\sigma_x$, $\sigma_y$ and $\sigma_z$ ). 

% While some authors include the diagonal Dirac $\mathbf{\beta}$ matrix inside the definition of the magnetisation operator \cite{gohr_phdthesis, spinDFT}, it is omitted in this formulation. The numerical differences resulting from this choice have been demonstrated to be negligible for chemical systems \cite{https://doi.org/10.1002/qua.22065}.
Standard approximated exchange-correlation energy functionals falling within the LDA or GGA frameworks are parametrized directly in terms of non-relativistic spin-polarized densities and their corresponding gradients:
\begin{eqnarray}
& E_{xc}^{LDA} = \int \epsilon_{xc}[\rho_+, \rho_-] \\ \nonumber
& E_{xc}^{GGA} = \int \epsilon_{xc}[\rho_+, \rho_-, \gamma_{++}, \gamma_{+-}, \gamma_{--}] .
\end{eqnarray}
where the gradient invariants are defined by:
\begin{equation}\gamma_{++} = \nabla \varrho_{+}\cdot \nabla \varrho_{+}, \quad\gamma_{+-} = \nabla \varrho_{+}\cdot \nabla \varrho_{-}, \quad\gamma_{--} = \nabla \varrho_{-}\cdot \nabla \varrho_{-}.\end{equation}
In a non-relativistic setting, $\varrho_{+}$ and $\varrho_{-}$ map uniquely to the spin-$\alpha$ and spin-$\beta$ electron densities. In the relativistic domain, however, explicit mappings for these  variables must be defined, leading to distinct collinear and non-collinear formalisms. In the density-only approximation there is not ambiguity, as the spin-polarized quantities are symmetrically mapped as $\varrho_{+}=\varrho_{-}=\frac{1}{2}\varrho$. 
For GGA functionals, this choice dictates that $\gamma_{++}=\gamma_{+-}=\gamma_{--}=\frac{1}{4}\nabla \varrho \cdot \nabla \varrho$. 

The exchange-correlation matrix elements are obtained by taking the functional derivative of the exchange-correlation 
energy with respect to the density matrix:
\begin{equation}
K_{\mu\nu}^{\text{TT}} = \frac{\delta E_{\text{xc}}}{\delta D_{\mu\nu}^{\text{TT}}}.
\end{equation}
For the  density-only approach ($\mathbf{m}(\mathbf{r})=0$), this derivative yields, for LDA:
\begin{equation}K^{\text{TT}}_{\mu\nu}=\int \left[\frac{\partial \epsilon_{xc}}{\partial \varrho_{+}}\chi^{\dagger}_{\mu}(\mathbf{r})\chi_{\nu}(\mathbf{r})\right]d{\mathbf{r}},\end{equation}
and for GGA:
\begin{equation}
K^{\text{TT}}_{\mu\nu}=\int \left[\frac{\partial \epsilon_{xc}}{\partial \varrho_{+}}\chi^{\dagger}_{\mu}(\mathbf{r})\chi_{\nu}(\mathbf{r})+\left(3\frac{\partial \epsilon_{xc}}{\partial \gamma_{++}}\nabla\varrho_{+}\right)\cdot\nabla\left(\chi^{\dagger}_{\mu}(\mathbf{r})\chi_{\nu}(\mathbf{r})\right)\right] d{\mathbf{r}}.\end{equation}
These density-only expressions will serve as numerical references for our collinear and non-collinear implementations in the limit for closed-shell systems. The derivation strategy used to obtain the explicit formulas for $K^{\text{TT}}_{\mu\nu}$ based on the multivariable chain rule is wholly general and was applied by Egidi et al. \cite{Egidi2017} and Petrone et al. \cite{Petrone2018}. Following their notation, the matrix elements $K^{\text{TT}}_{\mu \nu}$ can be obtained as:
\begin{equation}\label{eq:chainrule}
K_{\mu\nu}^{\text{TT}} = \frac{\delta E_{\text{xc}}}{\delta D_{\mu\nu}^{\text{TT}}} = \int \sum_{i,j} \frac{\delta \epsilon_{\text{xc}}}{\delta u_i} \frac{\delta u_i}{\delta V_j} \frac{\delta V_j}{\delta D_{\mu\nu}^{\text{TT}}} \, d\mathbf{r}.
\end{equation}
At this point we should clarify an important distinction regarding variable selection. While the fundamental independent variables of relativistic spin-density functional theory are the total charge density and the magnetisation vector field, most modern exchange-correlation functional libraries, such as Libxc \cite{Lehtola2018}, are designed around non-relativistic frameworks that evaluate energy densities ($\epsilon_{\text{xc}}$) explicitly using spin-polarised variables. The mathematical conversion between these two distinct representations is achieved seamlessly by the chain rule in Eq.~\eqref{eq:chainrule}, where $u_i$ denotes the internal spin-polarised functional inputs.
Specifically, $u_i \in \{\varrho_+, \varrho_-\}$ under the LDA framework, which expands to include the gradient invariants $\{\varrho_+, \varrho_-, \gamma_{++}, \gamma_{--}, \gamma_{+-}\}$ for GGA. $V_j$ denotes the physical variables specific to the chosen relativistic approximation. For example, in the collinear approach, these correspond to the total density and one magnetisation component, $V_j \in \{\varrho, m_z\}$, while in the non-collinear approach, they encompass the complete vector field, $V_j \in \{\varrho, m_x, m_y, m_z\}$ (or alternatively, $\{\varrho, |\mathbf{m}|\}$). We emphasise that identical matrix elements are obtained if the non-relativistic functionals are instead formulated in terms of the total density and the spin-polarisation density; for completeness this equivalence is explicitly demonstrated in Appendix~\ref{equivalence-framework}.

\subsection{Collinear (LDA and GGA) matrix elements}
The collinear approach closely mirrors non-relativistic theory
by utilizing only the $z$-component of the magnetisation vector to define the spin-polarized densities. 
This constraint  fixes the spin quantization axis along the  $z$-axis. Since the relativistic Hamiltonian does not commute with the spin operator, the collinear approach inherently violates rotational invariance \cite{van_wullen2002}. Consequently, molecular energies become unphysically dependent on their orientation in space. 
In the collinear framework, the spin-polarized densities are defined as follows:
\begin{equation}
    \rho_+(\mathbf{r}) = \frac{1}{2}[\rho(\mathbf{r}) + m_z(\mathbf{r})] \qquad \rho_-(\mathbf{r}) = \frac{1}{2}[\rho(\mathbf{r}) - m_z(\mathbf{r})]
\end{equation}
The same quantities can be rewritten as a function of the four-component spinors:
\begin{equation}
    \rho_\pm(\mathbf{r}) =  \frac{1}{2} \sum_i (\Psi_i^\dagger\Psi_i \pm \Psi_i^\dagger\mathbf{\Sigma}_z\Psi_i)
\end{equation}
and consequently substituted with the G-spinor basis set:
\begin{equation}\label{eq:pol_dens_collinear}
    \rho_\pm(\mathbf{r}) = \frac{1}{2}(\rho(\mathbf{r})  \pm m_z(\mathbf{r})) =\frac{1}{2}\sum_{T=L,S}\sum_{\mu\nu}\matel(\overlap{T} \pm \chi_\mu^{\dagger T}\sigma_z\chi_\nu^T)
\end{equation}
For LDA functionals, the exchange-correlation energy depends only on the local spin-polarised densities ($E_{\text{xc}}^{\text{LDA}} = \int \epsilon_{\text{xc}}[\varrho_+, \varrho_-]\,d\mathbf{r}$). Taking the functional derivative with respect to the density matrix element $D_{\mu\nu}^{\text{TT}}$ according to Eq.~\eqref{eq:pol_dens_collinear} yields the collinear LDA matrix elements:
\begin{eqnarray}
K_{\mu\nu \ (LDA)}^{TT} & = & \frac{\delta E_{xc}^{LDA}}{\delta \matel} = 
     \funcres{\rho_+} \frac{\partial \rho_+}{\partial \matel}\dr + \funcres{\rho_-} \frac{\partial \rho_-}{\partial \matel}\dr = \\ \nonumber
     & = & \frac{1}{2}  \funcres{\rho_+}[\overlap{T} + \chi_{\mu}^{\dagger T} \sigma_z \chi_{\nu}^{T}]\dr \ +   \frac{1}{2}\funcres{\rho_-}[\overlap{T} - \chi_{\mu}^{\dagger T} \sigma_z \chi_{\nu}^{T}]\dr.
\end{eqnarray}

The matrix elements for GGA functionals are  more complex because five 
partial derivatives must be evaluated, one for each internal variable upon which the GGA energy 
density depends ($\varrho_+, \varrho_-, \gamma_{++}, \gamma_{--}, \gamma_{+-}$):

%\begin{widetext} \label{eq:coll_gga}
\begin{eqnarray}
    K_{\mu\nu \ (GGA)}^{TT} && = \frac{\delta E_{xc}^{GGA}}{\delta \matel}
     = \funcres{\rho_+} \frac{\partial \rho_+}{\partial \matel}\dr + \funcres{\rho_-} \frac{\partial \rho_-}{\partial \matel}\dr + \\ \nonumber
     && + \funcres{\gamma_{++}} \frac{\partial \gamma_{++}}{\partial \matel}\dr + \funcres{\gamma_{--}} \frac{\partial \gamma_{--}}{\partial \matel}\dr + \funcres{\gamma_{+-}} \frac{\partial \gamma_{+-}}{\partial \matel}\dr = \\\nonumber
    && = \funcres{\rho_+} \frac{1}{2}[\overlap{T}  + \chi_{\mu}^{\dagger T} \sigma_z \chi_{\nu}^{T}]\dr + \\\nonumber
    && + \funcres{\rho_-} \frac{1}{2}[\overlap{T} - \chi_{\mu}^{\dagger T} \sigma_z \chi_{\nu}^{T}]\dr + \\\nonumber
     && + \funcres{\gamma_{++}} \{ \nabla\rho_+ \cdot\nabla[\overlap{T} + \chi_{\mu}^{\dagger T} \sigma_z \chi_{\nu}^{T}]\}\dr +\\\nonumber
     && + \funcres{\gamma_{--}} \{ \nabla\rho_- \cdot \nabla[\overlap{T} - \chi_{\mu}^{\dagger T} \sigma_z \chi_{\nu}^{T}]\}\dr + \\\nonumber
     && + \funcres{\gamma_{+-}} \{ \frac{1}{2}\nabla\rho_- \cdot \nabla[\overlap{T} + \chi_{\mu}^{\dagger T} \sigma_z \chi_{\nu}^{T}] + \\ \nonumber
     && + \frac{1}{2}\nabla\rho_+ \cdot \nabla[\overlap{T} - \chi_{\mu}^{\dagger T} \sigma_z \chi_{\nu}^{T}]\}\dr
\end{eqnarray}
These expressions require evaluating the  G-spinor overlap densities
$\chi_\mu^{\text{T}\dagger}(\mathbf{r})\chi_\nu^{\text{T}}(\mathbf{r})$ 
and the $z$ component of magnetisation overlap densities 
$\chi_{\mu}^{\text{T}\dagger}(\mathbf{r}) \sigma_z \chi_{\nu}^{\text{T}}(\mathbf{r})$ alongside their respective analytical gradients, $\nabla[\chi_\mu^{\text{T}\dagger}(\mathbf{r})\chi_\nu^{\text{T}}(\mathbf{r})]$ 
and $\nabla[\chi_{\mu}^{\text{T}\dagger}(\mathbf{r}) \sigma_z \chi_{\nu}^{\text{T}}(\mathbf{r})]$, ($\text{TT} = \text{LL}, \text{SS}$). The analytic gradient components of the G-spinors are evaluated using the procedures outlined in Ref.~\cite{Belpassi2005}.

\subsection{Canonical non-collinear matrix elements} 

The first attempt to go beyond the collinear approach is known as "canonical" \cite{van_wullen2002,JKubler_1988}.
Its development starts from the definition of a $2 \times 2$ generalized spin-density matrix $\mathbf{\overline{\varrho}}(\mathbf{r})$ \cite{PhysRevB.75.125119}:
\begin{eqnarray}
    && \overline{\rho} = \frac{1}{2}[\rho(\mathbf{r})\sigma_0 + \sum_c m_c(\mathbf{r})\sigma_c] = \\\nonumber
    && = \frac{1}{2}
\left [ \rho(\mathbf{r})
\left (
\begin{array}{cc} 1    &  0\\ 0    & 1\end{array}\right ) +
m_x(\mathbf{r})\left (\begin{array}{cc} 0   &  1\\ 1   & 0\end{array}\right ) +
m_y(\mathbf{r}) \left (\begin{array}{cc} 0   &  -i\\ i   & 0\end{array}\right ) + \right. \\\nonumber
&& \left. + m_z(\mathbf{r})\left (\begin{array}{cc} 1  &  0\\ 0   & -1\end{array}\right )
\right ] = \\
&& = \frac{1}{2} \left [ 
\begin{array}{cc}
 \rho(\mathbf{r}) + m_z (\mathbf{r})   & m_x(\mathbf{r})-im_y(\mathbf{r}) \\\nonumber
 m_x(\mathbf{r})+im_y(\mathbf{r})    & \rho(\mathbf{r}) - m_z (\mathbf{r})
\end{array}
\right]
\end{eqnarray}
where $c$ runs over $x,y,z$ and indicates the spatial components of the magnetisation vector. 
Diagonalizing this matrix at each point in space yields two local eigenvalues, which define the 
non-collinear local spin-polarized densities:
\begin{equation}
\label{eq:canonical_densities}
    \rho_+ = \frac{1}{2}[\rho(\mathbf{r}) + |{\bf m}|] \qquad\qquad \rho_- = \frac{1}{2}[\rho(\mathbf{r}) - |{\bf m}|]
\end{equation}
where $|{\bf m}|$ is the magnitude of the magnetisation vector, which will be indicated from now on as $m$ to simplify the notation:
\begin{equation}
    m = \sqrt{m_x^2 + m_y^2 + m_z^2}=\sqrt{\sum_cm_c^2}
\end{equation}
By avoiding an arbitrary reference frame, the local electron spin is quantized directly along the instantaneous orientation of the vector $\mathbf{m}(\mathbf{r})$. 
This ensures that the canonical approach strictly satisfies rotational invariance. 
Furthermore, since it accounts for all three spatial directions of the magnetisation vector
field inside the exchange-correlation functional, it recovers significantly more spin-correlation energy 
than its collinear counterpart.

By differentiating the LDA \textit{xc} functional with respect to the density matrix element
$D_{\mu\nu}^{\text{TT}}$ and employing the chain rule on the definitions in Eq.~\eqref{eq:canonical_densities}, the canonical LDA matrix elements are found to be:
%\begin{widetext}
\begin{eqnarray}\label{eq:nc_lda}
    K_{\mu\nu \ (LDA)}^{TT} && = \funcres{\rho_+} \frac{\partial}{\partial \matel}[\frac{1}{2}(\rho + m)]\dr + \funcres{\rho_-} \frac{\partial}{\partial \matel}[\frac{1}{2}(\rho - m)]\dr = \\\nonumber
    && = \funcres{\rho_+} \frac{1}{2}[\frac{\partial}{\partial \matel}\rho + \frac{\partial}{\partial \matel}m]\dr +  \funcres{\rho_-} \frac{1}{2}[\frac{\partial}{\partial \matel}\rho - \frac{\partial}{\partial \matel}m]\dr = \\\nonumber
    && = \funcres{\rho_+} \frac{1}{2}[\overlap{T} +\\\nonumber
    &&+ \sum_c\frac{m_c}{m}(\overlapsig{T})]\dr + \funcres{\rho_-} \frac{1}{2}[\overlap{T} - \sum_c\frac{m_c}{m}(\overlapsig{T})]\dr
\end{eqnarray}
%\end{widetext}
In the above expression we used:
\begin{equation}
\frac{\partial m}{\partial \matel}=\sum_c\frac{m_c}{m}(\overlapsig{T}).
\end{equation}
This formulation differs significantly from the collinear equivalent. The spinors are now coupled with  all three Pauli matrices ($\sum_c \sigma_c$) rather than being restricted to $\sigma_z$, and are modulated by the geometrical direction highlighting the absent choice of a preferential quantization axis.
% The presence of the magnetisation absolute value at the denominator could cause some numerical issues in the points of the grid where its value is very small. This does not directly affect the LDA functional \cite{Egidi2017}, but it will be crucial in the implementation of the GGA expressions.

 Extending the canonical formulation to GGA functionals yields a substantially more complex expression due to the inclusion of the spatial gradient of the magnetisation magnitude, $\nabla m(\mathbf{r}) = \sum_c \frac{m_c}{m}\nabla m_c$. Evaluating the total functional derivative leads to:
%\begin{widetext}
\begin{eqnarray}
     K_{\mu\nu \ (GGA)}^{TT} && = \funcres{\rho_+} \frac{1}{2}[(\overlap{T}) + \sum_c \frac{m_c}{m}(\overlapsig{T})]\dr + \funcres{\rho_-} \frac{1}{2}[(\overlap{T}) - \\ \nonumber
     && - \sum_c \frac{m_c}{m}(\overlapsig{T})]\dr + \\ \nonumber 
     && + \funcres{\gamma_{++}} \{ 2\nabla\rho_+ \cdot \frac{1}{2}\nabla[\overlap{T}] + (-\frac{1}{m^2}\sum_c \frac{m_c}{m}(\overlapsig{T})\sum_c m_c \nabla m_c + \\ \nonumber
    && + \frac{1}{m}\sum_c(\overlapsig{T})\nabla m_c + \frac{1}{m}\sum_c m_c \nabla[\overlapsig{T}])\}\dr + \\ \nonumber
    && + \funcres{\gamma_{--}} \{ 2\nabla\rho_- \cdot \frac{1}{2}\nabla[\overlap{T}] - (-\frac{1}{m^2}\sum_c \frac{m_c}{m}(\overlapsig{T})\sum_c m_c \nabla m_c + \\\nonumber
    && + \frac{1}{m}\sum_c(\overlapsig{T})\nabla m_c + \frac{1}{m}\sum_c m_c \nabla[\overlapsig{T}])\} \dr + \\ \nonumber
    && + \funcres{\gamma_{+-}} \{ \frac{1}{2}\nabla\rho_-\cdot[\nabla(\overlap{T}) + (-\frac{1}{m^2}\sum_c \frac{m_c}{m}(\overlapsig{T})\sum_c m_c \nabla m_c + \\ \nonumber
    && + \frac{1}{m}\sum_c(\overlapsig{T})\nabla m_c + \frac{1}{m}\sum_c m_c \nabla(\overlapsig{T}))] + \frac{1}{2}\nabla\rho_+ \cdot [\nabla(\overlap{T}) - \\ \nonumber
    && - (-\frac{1}{m^2}\sum_c \frac{m_c}{m}(\overlapsig{T})\sum_c m_c \nabla m_c  + \\\nonumber
    && + \frac{1}{m}\sum_c(\overlapsig{T})\nabla m_c + \frac{1}{m}\sum_c m_c \nabla(\overlapsig{T})]\}\dr
\end{eqnarray}
%\end{widetext}

Careful considerations must be made on the numerical issues of this expression. The presence in the denominator of the module of the magnetisation vector, which often assumes very small values (i.e. it is always zero for closed shell system), while not directly affecting the LDA functional \cite{Egidi2017}, is a potential source of instability for GGA. In the literature (see for instance Ref.\cite{crystal_2021}), it's common practice to drop the term originated by the gradient of $m_c/m$, meaning that two of the three terms derived from the gradient of the magnitude are approximated, and the expression then reads:
%\begin{widetext}
\begin{eqnarray}\label{eq:approx-canonical}
&& K_{\mu\nu \ GGA \ apx}^{TT} = \funcres{\rho_+} \frac{1}{2}[(\overlap{T} + \sum_c \frac{m_c}{m}(\overlapsig{T})]\dr + \funcres{\rho_-} \frac{1}{2}[(\overlap{T}) - \\\nonumber
&& - \sum_c \frac{m_c}{m}(\overlapsig{T})]\dr + \\\nonumber
&& +  \funcres{\gamma_{++}} \{ 2\nabla\rho_+ \cdot \frac{1}{2}(\nabla[\overlap{T}] + \frac{1}{m}\sum_c m_c \nabla[\overlapsig{T}])\}\dr + \\\nonumber
&& +  \funcres{\gamma_{--}} \{ 2\nabla\rho_- \cdot \frac{1}{2}(\nabla[\overlap{T}] - \frac{1}{m}\sum_c m_c \nabla[\overlapsig{T}])\}\dr + \\\nonumber
&& +  \funcres{\gamma_{+-}} \{\nabla\rho_- \cdot \frac{1}{2}(\nabla[\overlap{T}] + \frac{1}{m}\sum_c m_c \nabla[\overlapsig{T}])\}\dr + \\\nonumber
&& + \funcres{\gamma_{+-}} \{\nabla\rho_+ \cdot \frac{1}{2}(\nabla[\overlap{T}] - \frac{1}{m}\sum_c m_c \nabla[\overlapsig{T}])\}\dr.
\end{eqnarray}
%\end{widetext}

A similar approximate expression was reported in the context of DKS by Wang and Liu\cite{Wang2003_Polarization} and applied to open-shell atoms. The authors provide no details or discussion of issues concerning the numerical stability of this formulation. Although this truncation is expected to significantly suppress numerical grid noise, it remains an uncontrolled approximation, and the loss of exact physical gradient content is difficult to quantify. Even when using the simplified expression above, care must be taken to ensure numerical stability.
For example, in the CRYSTAL implementation \cite{crystal_2021}, despite the use of Relativistic Effective Core Potentials (RECPs), a multi-case thresholding approach has been applied for vanishing module magnetisation: i) when all three components $m_c$ fall below a critical global threshold, the singular ratio $m_c/m$ is regularised by averaging its value from the preceding self-consistent field (SCF) iteration over the local atomic Voronoi basin; ii) if a single spatial component dominates the magnetisation vector (i.e., $|m_z| \gg |m_x|, |m_y|$), the transverse contributions are locally truncated to zero and the collinear formulae are used at that specific grid point, aligning the local quantisation axis with the dominant component.

\subsubsection{Scalmani-Frisch formulation}\label{sec:sf}
The non-collinear framework developed by Scalmani and Frisch \cite{Scalmani2012} has been presented as an alternative formulation of non-collinear DFT. While the canonical method is notorious for generating severe numerical instabilities on the density integration grid, the Scalmani-Frisch formalism introduces a clever redefinition of the GGA variables to mitigate this behaviour.
Today, it stands as the most widely implemented non-collinear method, employed successfully at both the two-component \cite{Egidi2017} and four-component \cite{10.1063/1.5121713} relativistic levels. Since the method alters only the definition of the GGA variables depending from the gradient ($\gamma_{\sigma\sigma'}$), it is fully identical to the canonical approach in its LDA formulation.
Before proceeding, it is necessary to specify the notation introduced by Scalmani-Frisch to represent the scalar products: the $\cdot$ (dot) represents the scalar product between the components of the gradient ($\nabla\rho \cdot \nabla\rho = \sum_i \frac{\partial \rho}{\partial i}\frac{\partial \rho}{\partial i}$), while the $\circ$ (circle) symbol represents the scalar product between components of the magnetisation, so, by combining them, one can write:
\begin{equation}
    \nabla\mathbf{m}\cdot\circ\nabla\mathbf{m} = \sum_{c=x,y,z} \nabla m_c \cdot \nabla m_c
\end{equation}
where the sum over $c=x,y,z$ refers to the Cartesian components of the magnetisation vector and substitutes $\circ$, so the scalar product is now on the spatial components of the gradient.
Under this notation, the redefined GGA variables are expressed as:
\begin{equation}\label{eq:def-gamma-scalmani}
    \gamma_{\pm\pm} = \frac{1}{4}\nabla\rho \cdot \nabla\rho + \frac{1}{4}\nabla\mathbf{m}\cdot\circ\nabla\mathbf{m} \pm \frac{f_{\nabla}}{2}\sqrt{(\nabla\rho \cdot \nabla\mathbf{m}) \circ (\nabla\mathbf{m} \cdot \nabla\rho)}
\end{equation}

\begin{equation}
    \gamma_{+-} = \frac{1}{4}\nabla\rho \cdot \nabla\rho - \frac{1}{4}\nabla\mathbf{m}\cdot\circ\nabla\mathbf{m}
\end{equation}

\begin{equation}
    f_{\nabla} = sgn(\nabla\rho \cdot \nabla\mathbf{m} \circ \mathbf{m})
\end{equation}
where the $sgn$ function behaves as follows:
\begin{equation}
    sgn(x) = 
    \left\{
    \begin{array}{rl}
         +1 \  & \text{if} \ x>0 \\
         -1 \  & \text{if} \ x<0\\
         0 \  & \text{if} \ x=0
    \end{array}
        \right.
\end{equation}

Differentiating the GGA exchange-correlation energy  with respect to the density matrix elements $D_{\mu\nu}^{\text{TT}}$ within the Scalmani-Frisch GGA framework yields the following expression for the matrix elements:
%\begin{widetext}
\begin{eqnarray} \label{eq:scalmani}
    K_{\mu\nu \ GGA}^{TT} && = \funcres{\rho_+} \{ \frac{1}{2}[(\overlap{L}) + \sum_c \frac{m_c}{m}(\overlapsig{L})] \} + \\\nonumber
    && +\funcres{\rho_-} \{ \frac{1}{2}[(\overlap{L}) - \sum_c \frac{m_c}{m}(\overlapsig{L})] \} + \\ \nonumber
    && + \funcres{\gamma_{\pm\pm}} \{ [\frac{1}{2} \nabla\rho \cdot \nabla(\chi_{\mu}^{\dagger L}\chi_{\nu}^L)] + [\frac{1}{2}\sum_c \ \nabla m_c \cdot \nabla(\chi_{\mu}^{\dagger L}\sigma_c\chi_{\nu}^L)] \pm \\ \nonumber
    && \pm [\frac{\partial}{\partial \matel}[\frac{f_{\nabla}}{2}]\sqrt{(\nabla\rho \cdot \nabla\mathbf{m}) \circ (\nabla\mathbf{m} \cdot \nabla\rho)}] + \\ \nonumber
    && + \frac{f_\nabla}{2\bigstar}\sum_c (\nabla\rho \cdot \nabla m_c)[\nabla(\overlap{T})\cdot \nabla m_c + \nabla\rho \cdot \nabla(\overlapsig{T})]\} + \\\nonumber
    && + \funcres{\gamma_{+-}} \{ [\frac{1}{2} \nabla\rho \cdot \nabla(\chi_{\mu}^{\dag L}\chi_{\nu}^L)] + [\frac{1}{2}\sum_c \ \nabla m_c \cdot \nabla(\chi_{\mu}^{\dag L}\sigma_c\chi_{\nu}^L)] \}
\end{eqnarray}
%\end{widetext}
where the $\bigstar$ (star) symbol is a shorthand for  $\sqrt{(\nabla\rho \cdot \nabla\mathbf{m}) \circ (\nabla\mathbf{m} \cdot \nabla\rho)}$.

Some observations can be made regarding the new variables to clarify the reasoning behind their definition. First, the introduction of the \textit{signum} ensures strict consistency between the signs of the spin densities ($\varrho_+, \varrho_-$) and the gradient invariants when reducing to the collinear limit. Second, constructing the invariants using the vector field components $\mathbf{m}$ rather than its scalar magnitude $m$ is a direct attempt to avoid the numerical instabilities of the canonical approach.

However, although $m$ does not appear explicitly, the vector still appears in the denominator (as seen in Eq. \ref{eq:scalmani}), which is why sophisticated numerical  techniques are used to avoid singularities when the vector's value is too low, in both two-component \cite{Egidi2017, crystal_2021} and four-component implementations \cite{10.1063/1.5121713}.

% To clearly see this improvement, we can compare the definition of $\gamma_{++}$ in the canonical approach with its Scalmani-Frisch counterpart. In the canonical framework, $\gamma_{++}$ is defined as:
% \begin{eqnarray}
%     \gamma_{++} & = \nabla\rho_+ \cdot \nabla\rho_+ = (\frac{1}{2} \nabla\rho + \frac{1}{2}\nabla m) \cdot (\frac{1}{2} \nabla\rho + \frac{1}{2}\nabla m) = \\ \nonumber
%     & = \frac{1}{4} \nabla\rho \cdot \nabla\rho + \frac{1}{4}\nabla m \cdot\nabla m + \frac{1}{2}\nabla\rho \cdot \nabla m
% \end{eqnarray}

% Comparing this expression to Eq.~\eqref{eq:def-gamma-scalmani} reveals that the problematic squared magnitude gradient $\nabla m \cdot\nabla m$ is replaced by $\nabla\mathbf{m}\cdot\circ\nabla\mathbf{m}$.
% Although these two expressions are strictly related because $m=\sqrt{\mathbf{m}\circ\mathbf{m}}$, 
% the Scalmani-Frisch formulation completely eliminates the explicit use of the module of the magnetisation
% in the gradient.
The final cross-term does not map as straightforwardly, but it functions 
effectively as a highly stable mixed scalar product between the total density gradient and the magnetisation. 
% components, regularized via the $f_\nabla$ factor. 
%It must be noted, however, that while $m$ is 
% successfully removed from the definition of the $\gamma$ variables, it still appears in the denominator of the 
% underlying LDA terms inside Eq.~\eqref{eq:scalmani}. For this reason, spatial thresholding and regularization
% procedures remain essential in regions of vanishing magnetisation for both two-component \cite{Egidi2017, crystal_2021}
% and four-component implementations \cite{10.1063/1.5121713}.
Before moving on to describe our preliminary numerical results on the implementation of collinear and non-collinear LDA
scheme in the BERTHA code, in the following section we address the important issue of some exact conditions that  exchange-correlation functional should satisfy in the context of spin-DFT.

\subsection{Exact constraints of the exchange-correlation functionals: local and global torque theorems}
The exchange-correlation ($xc$)  functionals utilized in spin-polarized DFT (spin-DFT) 
are subject to rigorous exact constraints that must be preserved under any non-collinear redefinition of the spin variables. These constraints are formally established by the local and global torque theorems, originally derived by
Capelle and Vignale within non-relativistic time-dependent spin-DFT (TD-SDFT) \cite{capelle-vignale}. 
Despite their time-dependent origin, these theorems dictate general, fundamental conditions that any spin-polarized
functional must satisfy to properly capture the physical content of the system.
The first condition is the non-vanishing local torque theorem, which states that the local torque, defined as the vector product between the magnetisation density $\mathbf{m}(\mathbf{r},t)$ and the exchange-correlation effective magnetic field $\mathbf{B}_{\text{xc}}(\mathbf{r},t)$, cannot identically vanish:
\begin{equation} \label{eq:cap-vig}\mathbf{m}(\mathbf{r},t) \times \mathbf{B}{\text{xc}}(\mathbf{r},t) \neq \mathbf{0}.
\end{equation}
This theorem implies that $\mathbf{B}_{xc}(\mathbf{r},t)$, defined as:
\begin{equation}
    \mathbf{B}_{xc}(\mathbf{r},t)=\frac{\delta E_{xc}}{\delta \bf m}
\end{equation}
must not be parallel to the local magnetisation vector at any given point in space. A functional that violates this theorem remains entirely insensitive to spatial variations in the direction of the magnetisation vector field, as its energy expressions respond only to changes in the longitudinal component, that
is the module of the magnetisation. 
Furthermore, a vanishing local torque means that the \textit{xc} magnetic field cannot contribute directly to the spin dynamics of the system.
Integrating Eq.~\eqref{eq:cap-vig} over all space yields the second exact constraint:
\begin{equation}
\int \mathbf{m}(\mathbf{r},t) \times \mathbf{B}_{xc}(\mathbf{r},t) d\mathbf{r} = \mathbf{0}.
\end{equation}
This condition requires that the internal, self-consistent exchange-correlation magnetic field cannot exert a net macroscopic torque on the system as a whole. Preserving these constraints is essential for any method that aims to reliably simulate explicit real-time spin dynamics. While the global zero-torque theorem is easily satisfied by most standard non-collinear approximations, satisfying the non-vanishing local torque theorem is notoriously difficult. For example, the traditional canonical non-collinear approach \cite{PhysRevB.75.125119} fails this local condition because its effective $\mathbf{B}_{xc}$ field is, by definition, forced to align parallel to $\mathbf{m}(\mathbf{r})$ at every grid point. In contrast, the Scalmani-Frisch formulation \cite{Scalmani2012} satisfies the theorem by incorporating the transverse gradient signatures directly into the definition of the $\gamma_{\pm\pm}$ invariants.
The proofs for these theorems are provided in Appendix \ref{proof_torque} for both the canonical non-collinear and the Scalmani-Frisch approaches, using a common notation.
Beyond these two theorems, Vignale and Rasolt \cite{vignale2} demonstrated that a complete description of spin dynamics requires an asymmetric formulation that treats the transverse and longitudinal gradients of the magnetisation vector field differently. To date, this highly stringent condition has been successfully satisfied only by the multicollinear approach \cite{multicollinear}; however, detailed numerical comparisons between different approaches are lacking in the literature.

\section{Results}

The following results present the performance of the implemented collinear and non-collinear DKS schemes across a series of open-shell hydride molecules, focusing on the effects of increasing spin-orbit coupling. Firstly we analyze the breakdown of energy rotational invariance in collinear calculations, then follow by an evaluation of how our reformulated non-collinear canonical approach restores this invariance and facilitates a robust description of molecular dissociation

\subsection{Computational details}

As previously outlined, this initial work, in addition to deriving the formulas for the collinear and non-collinear canonical and Scalmani-Frisch approaches, presents a four-component implementation of the collinear method for both LDA and GGA functionals, alongside the non-collinear canonical scheme restricted to the LDA formulation. Since these specific approaches avoid the highly divergent magnetisation gradient terms characteristic of non-collinear gradient corrections, they are expected to remain numerically stable.
The extension to non-collinear GGA functionals, which introduces significant numerical challenges on the integration grid, is postponed to future work. The implementation has been incorporated into the DKS module of the BERTHA electronic structure package \cite{10.1063/5.0002831} and added to a dedicated repository \cite{githubBERTHA4cDKS}. As the current subroutines evaluate the exchange-correlation contributions to the DKS matrix elements by explicitly calculating G-spinor amplitudes (and their gradients) without auxiliary density-fitting techniques, the associated computational cost is relatively high. Consequently, the numerical benchmarks presented here focus on small, representative molecular systems containing at most a single heavy atom. The code parallelisation and implementation of a density-fitting algorithm are highly desirable and are currently in progress to extend the practical applicability of these relativistic methods to larger chemical systems.

To analyse the effect of spin-orbit coupling on the magnetisation vector field, we evaluated both the Cartesian components of the magnetisation vector ($\bf m$) and its modulus ($m$), and developed a basi graphical tools to investigate its local behaviour by visualising the local magnetisation density ($\bf m(r)$) and its ($ m({\bf r})$).
\begin{align}
{m}_c(\mathbf{r}) = \sum_i \mathbf{\Psi}_i^\dagger(\mathbf{r}){\Sigma}_c\mathbf{\Psi}_i(\mathbf{r}) \quad &\Longrightarrow \quad m_c = \int \left[ \sum_i \mathbf{\Psi}_i^\dagger(\mathbf{r})\mathbf{\Sigma}_c\mathbf{\Psi}_i(\mathbf{r}) \right] d\mathbf{r}, \\
m(\mathbf{r}) = \sqrt{\sum_c m_c^2(\mathbf{r})} \quad &\Longrightarrow \quad m=\int \sqrt{\sum_c m_c^2(\mathbf{r})}  d\mathbf{r},
\end{align}
where $c \in \{x, y, z\}$. These properties are mapped onto 3D volumetric isosurfaces to display individual directional intensities, and 2D vector maps across the $zx$-plane to show local orientations.
%To diagnose the numerical instabilities inherent to the canonical approach, variables are evaluated and compared directly along the internuclear axis of diatomic test cases.

To validate the four-component DKS implementation, initial benchmark calculations were performed using the density-only and collinear frameworks to assess numerical precision and determine the number of reliable significant digits. 
%These reference calculations were carried out using the exact two-component (X2C) approximation %\cite{10.1063/1.1413512} and the Zeroth-Order Regular Approximation (ZORA) \cite{10.1063/1.472460}, %respectively. 
Reference calculations have been carried out using restricted and unrestricted non-relativistic KS calculations with the ORCA \cite{ORCA5} and PySCF \cite{PYSCF} packages. Within the BERTHA framework, the non-relativistic limit is evaluated by scaling the speed of light by a factor of 100 ($100c$), effectively eliminating most relativistic kinematic corrections and spin-orbit coupling. The geometries of the molecular test sets were optimised under strict $C_{2v}$ spatial symmetry using the ORCA package in conjunction with the relativistic ZORA Hamiltonian for closed-shell molecules; the resulting coordinates are provided in the Supporting Information (SI). All calculations were performed using uncontracted dyall.v2z basis sets \cite{dyall2023b}, retrieved directly from the Basis Set Exchange repository \cite{pritchard2019a}, while real-space numerical integration grids were maintained at their respective software defaults. Density fitting and resolution-of-identity techniques were explicitly deactivated for all runs. As for the exchange-correlation functionals, the Slater local density approximation \cite{Dirac1930_376} was chosen for the LDA functionals, while the Perdew-Burke-Ernzerhof (PBE) formulation\cite{pbe} was selected to represent the GGA class, with both functionals evaluated via the standardised Libxc exchange-correlation library \cite{Lehtola2018}.

\subsection{Collinear results: LDA and GGA}

As an initial test of the  correctness of our new collinear implementation, 
we verified that the reformulated equations yield the exact density-only limit for closed-shell systems.
The resulting numerical data for the H$_2$O molecule are presented in Table S1 of the Supporting Information (SI).
The benchmarks demonstrate that the collinear approach reproduces the total energy of the density-only formulation up to eleven significant digits, a level of agreement perfectly compatible with the inherent numerical noise of the real-space integration grid employed in BERTHA.

To further evaluate the robustness of our implementation, we investigated the performance of the new collinear scheme on simple open-shell molecular systems. Since the collinear DKS scheme must formally reduce to the standard unrestricted Kohn-Sham (UKS) model in the non-relativistic limit the approach was benchmarked with results obtained with ORCA and PySCF.
%we benchmarked our implementation against reference data obtained from two widely adopted quantum chemistry software packages, PySCF \cite{PYSCF} and ORCA \cite{ORCA5}.
Absolute total energies for both the closed-shell water molecule and its corresponding open-shell radical cation ($\text{H}_2\text{O}^+$) are reported in Table S2-S3 of the SI. A comparative analysis of these numerical data demonstrates highly consistent values across all three implementations. Specifically, the calculated vertical ionisation energies vary by less than $0.0002$ and $0.001 \text{ eV}$ with respect to PySCF and ORCA, confirming that the four-component collinear implementation accurately recovers the non-relativistic UKS limit with negligible numerical deviation (the slight difference can be attributed to the different numerical integration schemes).

To validate the implementation under pronounced relativistic conditions, where the inclusion of SOC is expected to break rotational invariance, our results were benchmarked against data available in the literature. Desmarais et al. \cite{crystal_2021} previously evaluated this effect by incorporating the SOC operator self-consistently within a relativistic two-component framework.
In their study, the authors evaluate the loss of rotational invariance in the total energy of the open-shell, linear $\text{I}_2^+$ molecule by comparing collinear and non-collinear approaches across different spatial orientations.
The absolute total energies, evaluated at three distinct spatial orientations ($0^\circ$, $45^\circ$, and $90^\circ$), are summarized in Table \ref{tab:i2plus_confr}. Taking the $0^\circ$ geometry as reference, our collinear DKS implementation reproduces the order of magnitude of the energy difference observed in the literature, which remains consistently on the order of $1\cdot10^{-3}\text{ a.u.}$ (about 0.027 eV). Since the values calculated with the CRYSTAL code were obtained using different computational parameters, most notably the employment of relativistic effective core potentials (pseudopotentials) rather than our all-electron four-component Dirac spinors, an exact numerical match cannot be expected.
\begin{table}[!h]
    \centering
    \begin{tabular}{cccc}
    \toprule
    Angle & {Energy (a.u.)} & {Difference} & {Crystal \cite{crystal_2021}} \\
    \midrule
       0°  & -14237.306588 & 0 & {0} \\
       45° & -14237.304350 & {$2.2 \times 10^{-3}$} & {$1.5 \times 10^{-3}$} \\
       90° & -14237.304728 & {$1.9 \times 10^{-3}$} & {$1.6 \times 10^{-3}$} \\
       \bottomrule
    \end{tabular}           
    \caption{Total energy (a.u.) of $\mathrm{I_2^+}$ in different geometry arrangements (0°,45°,90°). The energy difference calculated by taking as a reference the 0° geometry is also reported and compared with the results in literature.}
    \label{tab:i2plus_confr}
\end{table}

\begin{figure}[htbp]
    \centering
    \includegraphics[width=0.75\textwidth]{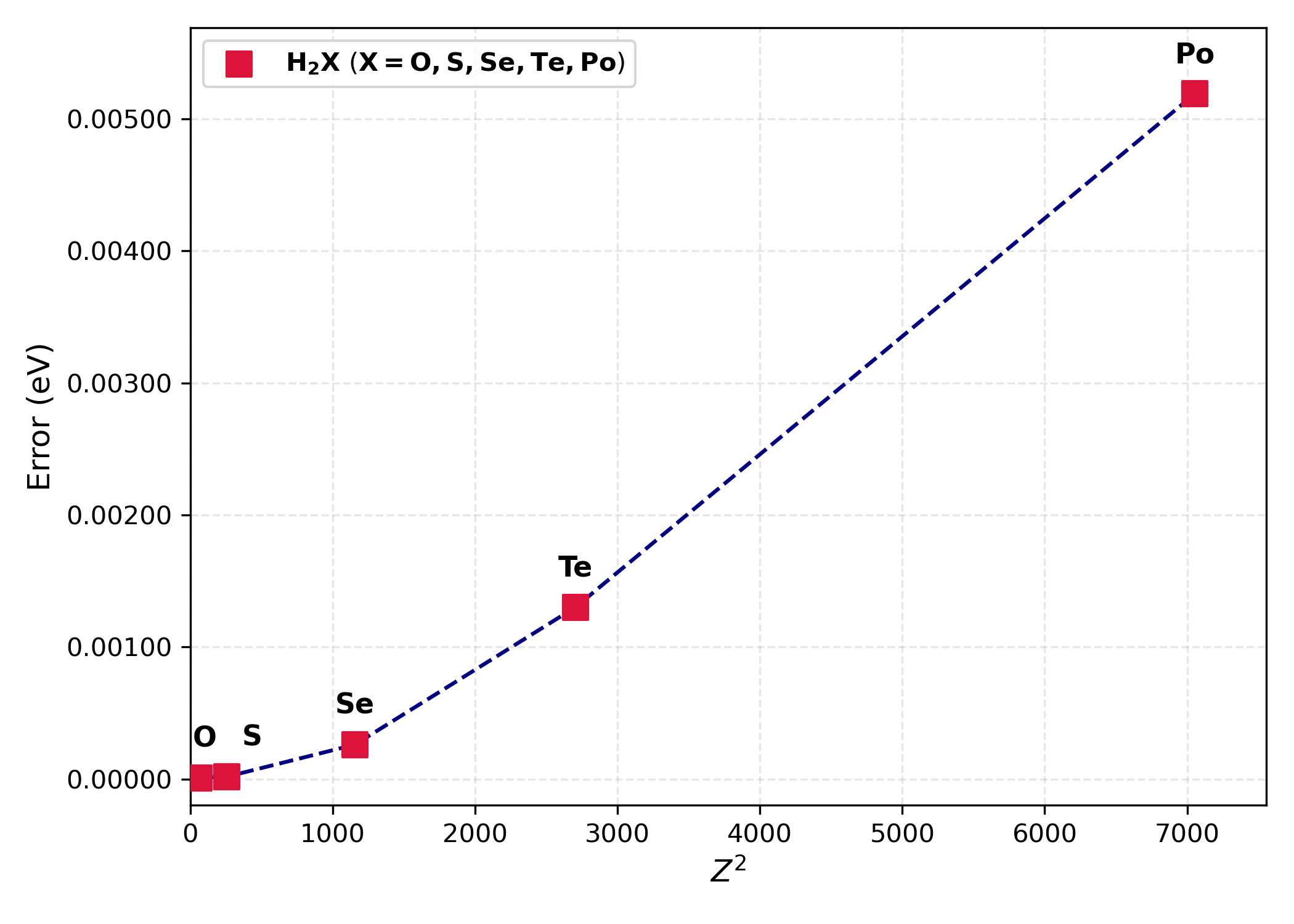}
    \caption{Breaking of rotational invariance in the collinear implementation in \texttt{BERTHA}. The absolute difference in vertical ionization energy between the 0° and 90° configurations is plotted as a function of the square of the chalcogen atomic number ($Z^2$) for the complete $\mathrm{H_2X}$ ($\mathrm{X = O, S, Se, Te, Po}$) homologue series.}
    \label{fig:orientation_error_linear}
\end{figure}

Next, we applied our BERTHA implementation of the collinear approach to a series of simple triatomic molecules to investigate how the breaking of rotational invariance manifests in practice. 
To this aim, we selected the chalcogen hydride series, $\mathrm{H_2X}$ ($\mathrm{X = O, S, Se, Te, Po}$), 
which exhibits a systematically increasing SOC effect. 
Each molecule has been calculated in two distinct spatial orientations: firstly with the molecular $C_2$ symmetry axis aligned along the Cartesian $z$-axis (designated as the 0 degree geometry), and then after a rigid space rotation that aligns the symmetry axis with the $x$-axis (the 90 degrees geometry).
The resulting vertical ionisation energies for both configurations are reported in Table \ref{tab:confr-en-diff}. Since the closed-shell neutral ground state energy is invariant under rotation, the observed variations arise exclusively from the orientation-dependent behaviour of the open-shell cationic species. As summarised in the Table, the discrepancy in total energy induced by the choice of the global magnetisation quantisation axis increases by three orders of magnitude moving down the chalcogen group, ranging from approximately $10^{-6}\,\text{eV}$ for $\text{H}_2\text{O}^+$ to approximately $10^{-3}\,\text{eV}$ for $\text{H}_2\text{Te}^+$ and $\text{H}_2\text{Po}^+$. This trend directly parallels the strengthening of the SOC effect, which remains negligible for light elements but becomes highly pronounced in heavy-atom systems (see Figure \ref{fig:orientation_error_linear}). Notably, for the heavier molecules where this energy difference becomes numerically more significant, the magnitude of the energy deviation increases approximately linearly with $Z^2$. This scaling behaviour is consistent with the physical nature of relativistic spin-orbit interactions, which are indeed expected to scale as $Z^2$.

\begin{table}[h!]
    \centering
    \begin{tabular}{cccccc}
    \toprule
   Angle & {$\mathrm{H_2O}$} & {$\mathrm{H_2S}$} & {$\mathrm{H_2Se}$} & {$\mathrm{H_2Te}$} & {$\mathrm{H_2Po}$} \\
   \midrule
   0 & 12.4349 & 10.3008 & 9.6489& 9.1092 & 8.5542 \\
  90 & 12.4349 & 10.3008 & 9.6486& 9.1079 & 8.5490 \\
    \midrule
    &  {$9\times10^{-6}$} & {$1.7\times10^{-5}$} & {$2.6\times10^{-4}$} & {$1.3\times10^{-3}$} & {$5.19\times10^{-3}$} \\
    \bottomrule
\end{tabular}
\caption{Ionization energies and their differences (in eV) calculated with collinear $z$ approach as implemented in BERTHA, at the relativistic four-component level, for $\mathrm{H_2X}$ (X= O, S, Se, Te, Po) at 0° and 90° orientations}
\label{tab:confr-en-diff}
\end{table}

\begin{figure}
    \centering
    \includegraphics[width=0.45\linewidth]{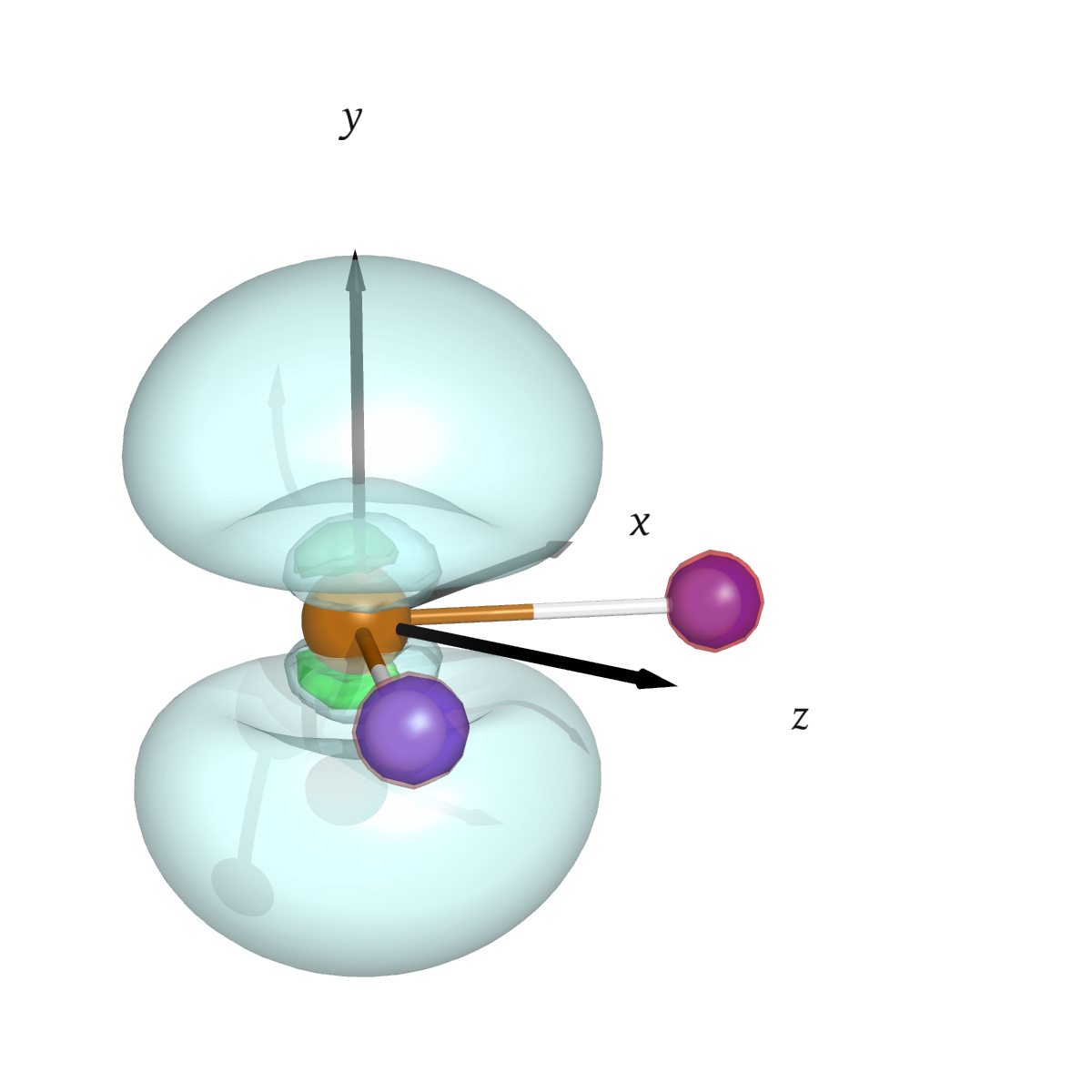}
    \includegraphics[width=0.45\linewidth]{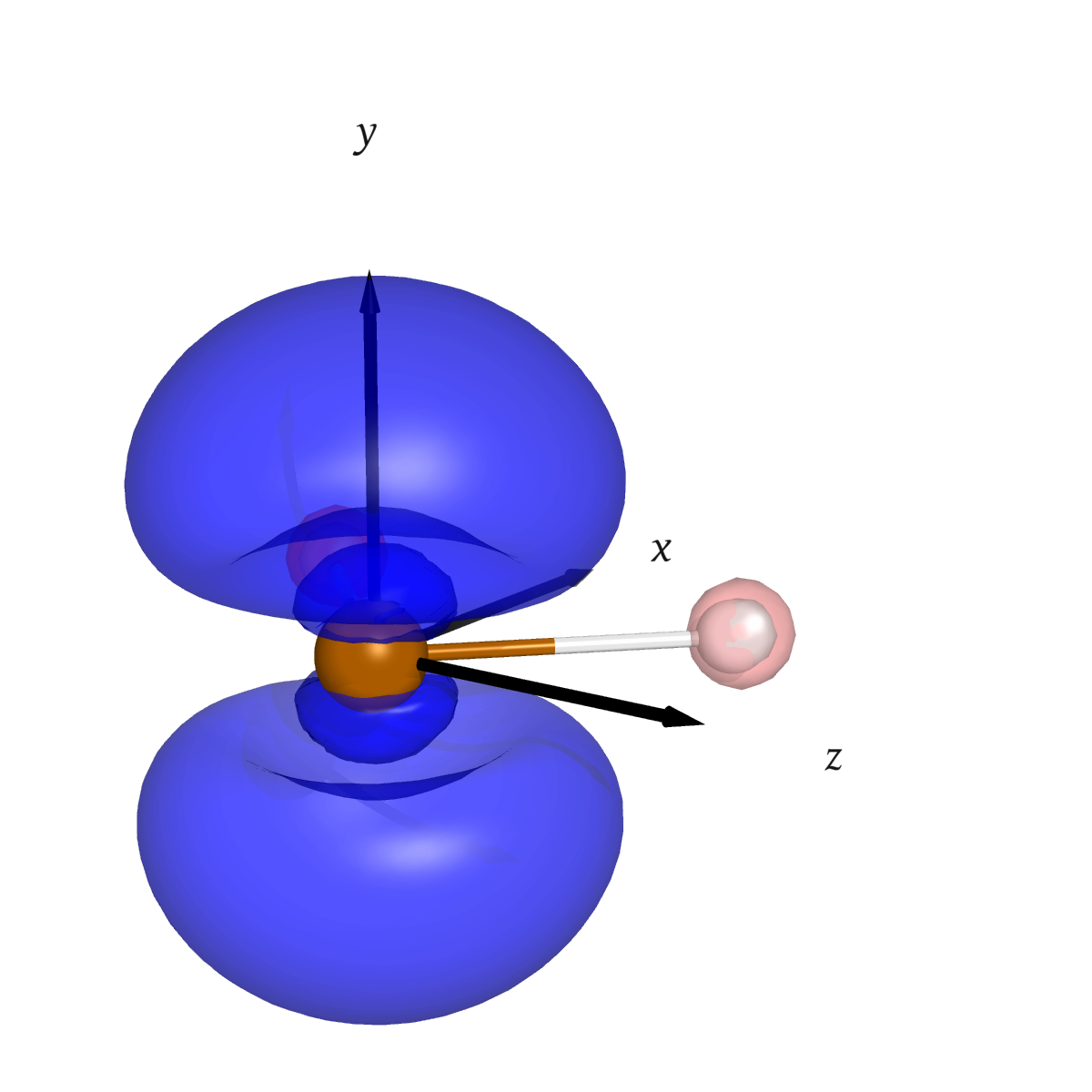}
    \caption{Isosurfaces of the magnetisation vector components for \mathtxt{H_2Po^+} at 0° (left) and 90° (right) geometry, calculated with collinear $z$ as implemented in BERTHA. Components \mathtxt{m_x}, \mathtxt{m_y} and \mathtxt{m_z} in red, green and blue, respectively. The different shades represent a change in sign. }
    \label{fig:magne_art_h2po}
\end{figure}

To gain deeper insight into the origin of non-rotational invariance, the module of the integrated magnetisation vector ($m$) and its Cartesian components ($m_x, m_y, m_z$) were evaluated. These values, calculated at the four-component relativistic level using the BERTHA code under a collinear $z$-axis constraint, are summarised in Table \ref{tab:mag_collinear}
for both the $0^\circ$ and $90^\circ$ geometries of $\text{H}_2\text{O}^+$ and $\text{H}_2\text{Po}^+$ 
(the complete spatially resolved integrated magnetisation data are compiled in Tables S4-S5 of the SI).
For $\text{H}_2\text{O}^+$, a system with negligible  SOC, the collinear framework yields an integrated $m_z$ component that is numerically identical to the total magnitude
$m$, confirming that the magnetisation aligns perfectly along the designated quantization axis. 
The total energy  remains invariant under a rigid spatial rotation, producing identical component distributions for both the $0^\circ$ and $90^\circ$ orientations. Conversely, for the heavy $\text{H}_2\text{Po}^+$ analogue, a significant non-vanishing contribution from the 
perpendicular $m_y$ component arises, meaning that $m_z$ is no longer the unique component 
defining the magnetisation vector. 
More importantly, the individual Cartesian contributions redistribute upon spatial rotation, indicating that the  spin-orbit coupling  modulates the relative components of the magnetisation density depending on the molecular orientation relative to the global quantization axis.
This phenomenon is visually demonstrated in Figure \ref{fig:magne_art_h2po}, which compares the magnetisation density component isosurfaces for $\text{H}_2\text{Po}^+$ at $0^\circ$ and 
$90^\circ$.
%(the complete set of isosurfaces for both $\text{H}_2\text{O}^+$ and $\text{H}_2\text{Po}^+$ is available in the SI, Figures S1-S2). 
A clear discrepancy in the spatial intensity distribution is observable between the two orientations, directly mirroring the integrated values. 
In particular, the left panel ($0^\circ$ configuration) in Figure \ref{fig:magne_art_h2po} exhibits a distinct green isosurface corresponding to the $m_y$ component, which
is absent in the right panel ($90^\circ$ configuration).
This orientation-dependent behaviour of the magnetisation components is at the basis of the breaking of rotational invariance within the collinear framework. 

\begin{figure}
    \centering
    \includegraphics[width=0.90\linewidth]{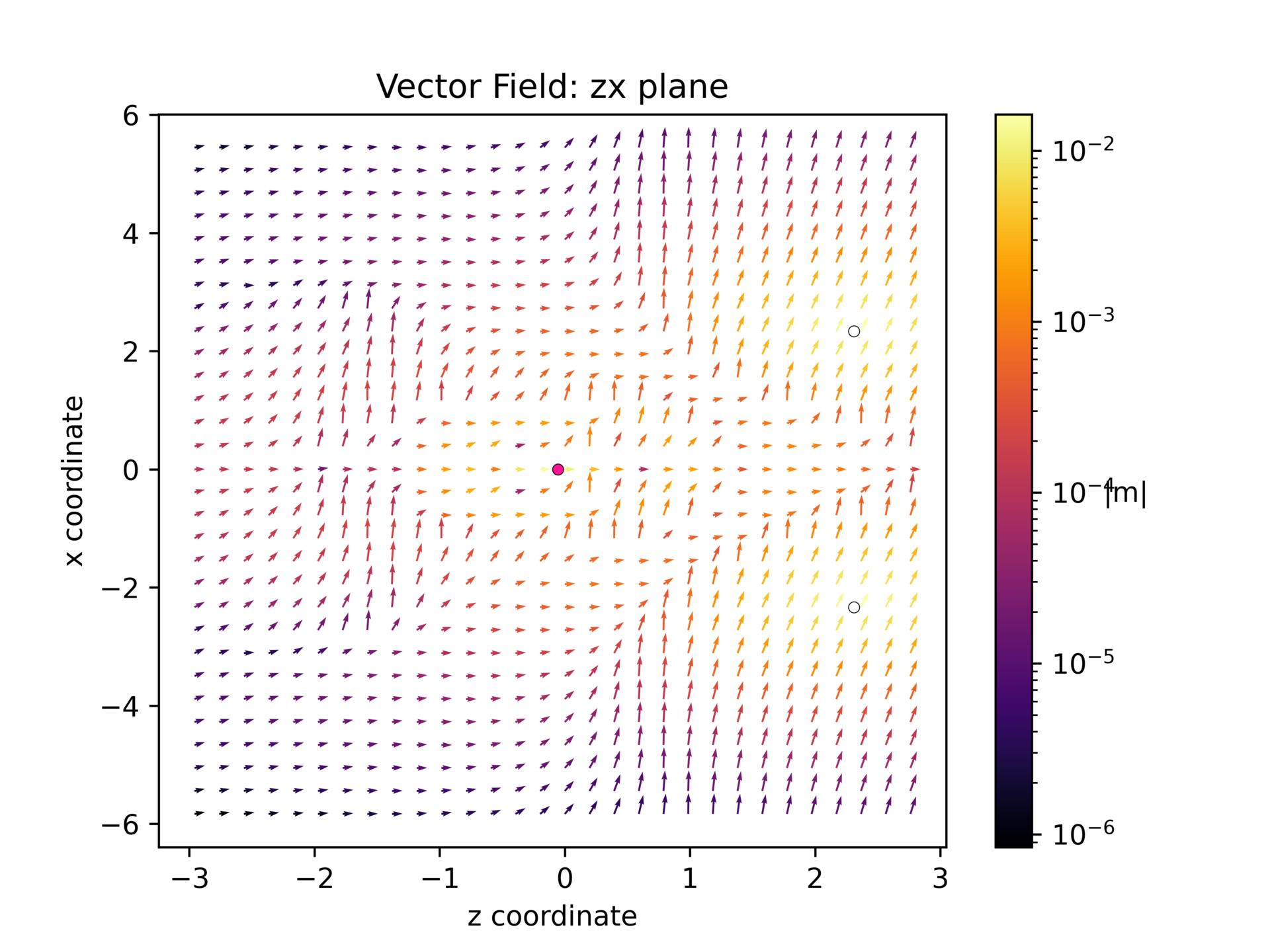}
    \includegraphics[width=0.90\linewidth]{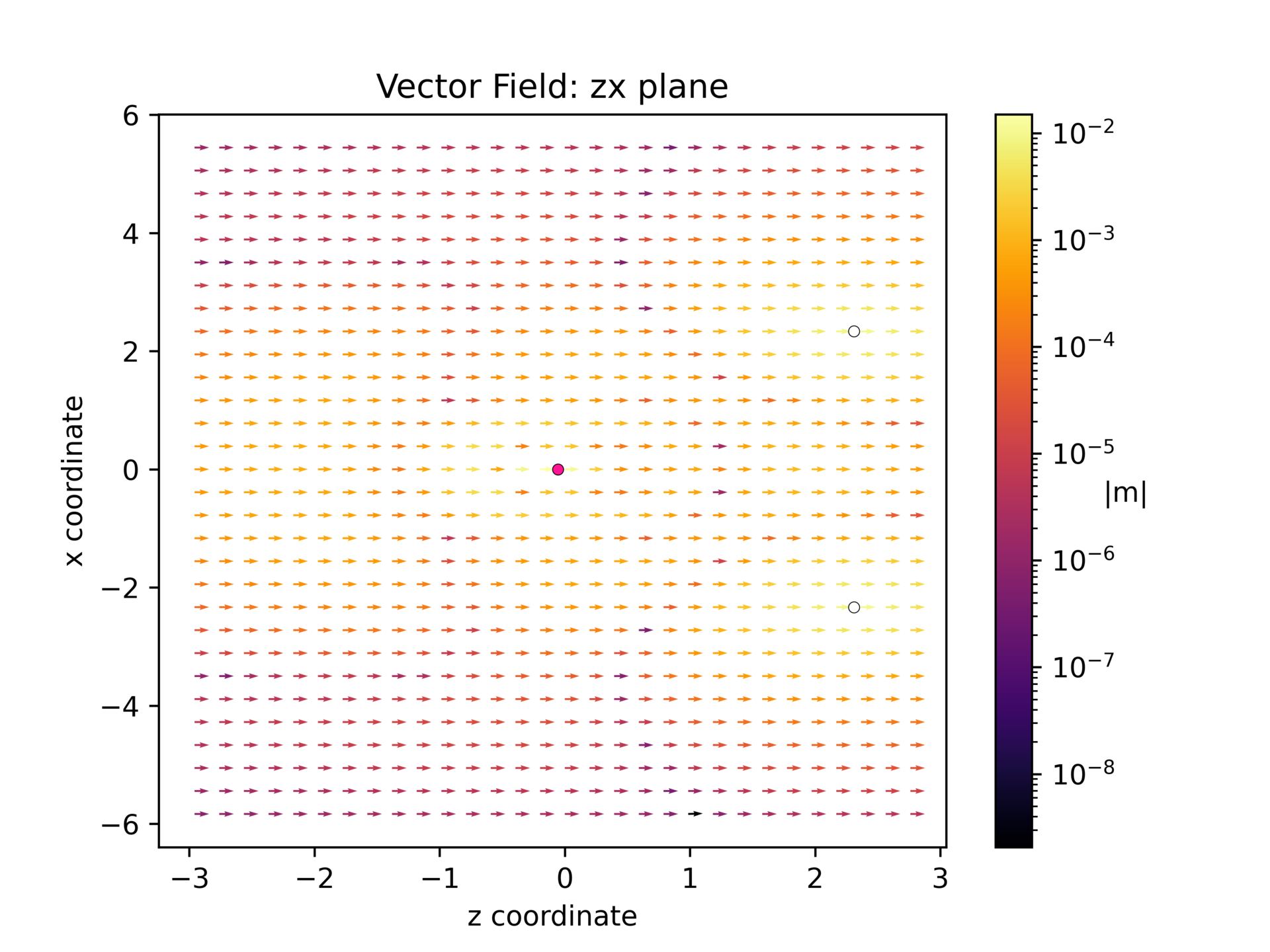}
    \caption{Vector map of the zx plane (y=0). The arrows represent the direction of the magnetisation vector, the colour its intensity. Top panel: normal relativistic effects; bottom panel: $c \rightarrow \infty$ limit}
    \label{fig:vector_c100_coll}
\end{figure}

A striking visual impact of the SOC effect is provided by the two-dimensional vector field plots mapping the orientation of the magnetisation vector within the $xz$ plane containing the molecule (Figure \ref{fig:vector_c100_coll}). In the presence of SOC, the spatial orientation of the magnetisation vector is position-dependent and varies continuously across the molecular domain, with non-vanishing integrated transverse components. Conversely, in the non-relativistic limit (obtained by increasing the speed of light, $c \rightarrow c \cdot 100$), the vector field exhibits perfectly uniform, unidirectional collinear alignment. This stark contrast highlights that a strictly collinear description is physically valid only when relativistic spin-orbit effects are negligible.
Consequently, the analysis of both integrated values and spatial density isosurfaces serves as a diagnostic tool for understanding the fundamental breakdown of the collinear approximation in systems governed by spin-orbit coupling. Within a fully relativistic framework, the components of the magnetisation vector are intrinsically linked to the spatial electronic structure, and their specific spatial configurations are implicitly determined by the SOC terms in the Hamiltonian. Since a standard collinear exchange-correlation functional depends solely on the longitudinal component ($m_z$), any contribution arising in the transverse directions ($m_x$ and/or $m_y$) is inherently neglected and does not contribute to the exchange-correlation energy or the associated potential.
\begin{table}[h!]
    \centering
    \begin{tabular}{ccccc}
    \toprule
    &  {\mathtxt{H_2O^+} (0°)} & {\mathtxt{H_2O^+} (90°)} &  {\mathtxt{H_2Po^+} (0°)} & {\mathtxt{H_2Po^+} (90°)} \\
    \midrule
    \mathtxt{m_x} & 1.099$\times 10^{-4}$ & 1.140$\times 10^{-4}$ & 8.724$\times 10^{-2}$ & 8.565$\times 10^{-2}$ \\
   \mathtxt{m_y} & 3.906$\times 10^{-3}$ & 1.444$\times 10^{-3}$ &0.205 & 9.872$\times 10^{-2}$ \\
   \mathtxt{m_z} & 1.061 & 1.061 &0.995 & 1.002 \\
   \mathtxt{|m|} & 1.061 & 1.061 &1.070 & 1.064\\
   \bottomrule
    \end{tabular}
    \caption{Values of integrated magnetisation vector components and absolute value for \mathtxt{H_2O^+} and \mathtxt{H_2Po^+}, calculated at the relativistic four-component level with BERTHA (collinear $z$). Both 0° and 90° geometries are reported.}
    \label{tab:mag_collinear}
\end{table}

To further explore the entity of the breaking of rotational invariance,
the linear coinage metal hydrides $\text{AgH}$ and $\text{AuH}$ were analyzed using the collinear $z$
approach. Results are summarized in Table \ref{tab:hydrides_diff}.
 For AgH, the difference between the ionisation energies calculated at the two spatial orientations is approximately $2\times10^{-4}$, which is an order of magnitude lower than that found for tellurium, although both atoms are in the same row of the periodic table (which would suggest an effect of similar magnitude). In contrast, the effect for gold is of the same order as that calculated for polonium ($\sim 5\times10^{-3}$).
This behaviour is not easy to rationalise, even in light of the different nature of the spinors involved in the ionisation process in the two cases. In $\text{AgH}^+$ and $\text{AuH}^+$, the unpaired electron resides in a molecular $\sigma$-orbital formed primarily by the isotropic valence $s$-orbitals ($\text{Ag } 5s$ / $\text{Au } 6s$) overlapping with $\text{H } 1s$. In contrast, the open-shell electron in the bent $C_{2v}$ chalcogenide cations occupies a highly directional valence molecular orbital with dominant $p$-type character.
Interestingly, van W\"{u}llen \cite{van_wullen2002} reports that if the electrons involved in the ionisation belong to an orbital with spherical symmetry, one would not observe the effects of rotational invariance breaking. While this has been established for atoms, it is clearly not obvious for molecules.
\begin{table}[h]
    \centering
    \begin{tabular}{cc|c|c|c}
    \hline
    & \multicolumn{2}{c}{\mathtxt{AgH}} &\multicolumn{2}{c}{\mathtxt{AuH}} \\
    \hline
     & {NR} & {R} & {NR} & {R}\\
    \hline
     0 ° & 9.0933 & 9.4650 & 9.0340 & 10.1666 \\
    90°  & 9.0933 & 9.4648 & 9.0335 & 10.1613 \\
    Difference & 0.0 & 0.0002 & 0.0005 & 0.0053 \\
    \hline
    \end{tabular}
    \caption{Ionization energies and their differences (in eV) calculated with the collinear $z$ approach as implemented in BERTHA, at relativistic (R) and non-relativistic (NR) level, for AgH and AuH at two different orientations (0° and 90°)}
    \label{tab:hydrides_diff}
\end{table}

\subsection{Non-collinear LDA}
In this section, we present some results obtained using the non-collinear LDA formulation outlined in Eq. (\ref{eq:nc_lda}). 
We focus specifically on the restoration of rotational invariance, 
a property expected from the canonical non-collinear approach even at the local density approximation. 
This restoration arises from the fact that the canonical framework depends strictly on the scalar magnitude of the local magnetisation vector field, which is inherently invariant under rigid spatial rotations of the molecular system. Conversely, extending this treatment to GGAs and implementing the complex Scalmani-Frisch ansatz, both of which require meticulous algorithmic care to preserve numerical stability, constitute a demanding undertaking that will be the subject of a forthcoming work. We note that all results have been obtained without applying any screening technique in Eq. \ref{eq:nc_lda}.

As a preliminary test, the \mathtxt{H_2O} molecule was analysed using a reduced value of the speed of light (c = c/8) to artificially enhance relativistic effects and allow observation of the influence of SOC even in light molecules.
An illustrative comparison of collinear and canonical LDA results for \mathtxt{H_2O} is presented in Table \ref{tab:canonical_lda}. It is immediately apparent that, while the collinear approach yields a difference of \mathtxt{10^{-2}} eV between the two ionisation energies, this difference is reduced by three orders of magnitude when a non-collinear method is applied, demonstrating restoration of rotational invariance up to 10 significant digits in the open-shell energy.
\begin{table}[h!]
    \centering
    \begin{tabular}{ccc}
    \toprule
    Angle & Collinear & Canonical \\
    \midrule
    0°  & 10.78989 & 10.74725 \\
    90° & 10.75693 & 10.74727 \\
    \midrule
    Difference & 0.03296 & 0.00002 \\
    \bottomrule
    \end{tabular}
    \caption{Ionization energy  (eV) of \mathtxt{H_2O}, calculated with collinear $z$ and canonical non-collinear LDA approaches as implemented in BERTHA at 0° and 90° geometries. Calculations carried out with a  reduced  speed of light ($c=c/8$).}
    \label{tab:canonical_lda}
\end{table}

In Table \ref{tab:iodio-polonio} the results of the analysis carried out with the canonical non-collinear LDA implementation  on the energies of \mathtxt{I_2^+} and \mathtxt{H_2Po^+} molecules are reported, for which a significant SOC effect is expected based on the previous analysis with the collinear implementation.
The rotational invariance is in both cases $\sim 10^{-5}$a.u. reaching a 
numerical rotational invariance on the total energy up to ten significant digits.
\begin{table}[h!]
    \centering
    \begin{tabular}{ccc}
    \toprule
    Angle & \mathtxt{I_2^+} & \mathtxt{H_2Po^+} \\
    \midrule
   0°      & -14215.589960 &  -22220.733429   \\
   90°     & -14215.590053 &  -22220.733419   \\
   \midrule
    Difference & -9.3 $\times 10^{-5}$ & 1.0 $\times 10^{-5}$ \\
    \bottomrule
    \end{tabular}
    \caption{Energies (a.u.) of \mathtxt{I_2^+} and \mathtxt{H_2Po^+} at 0° and 90° geometry, carried out with canonical LDA as implemented in BERTHA.}
    \label{tab:iodio-polonio}
\end{table}

It is instructive to analyze the integrated values and spatial isosurfaces of the 
magnetisation density components. The results reported in Table \ref{tab:mag_canonical_lda}
for the $\text{H}_2\text{O}^+$ and $\text{H}_2\text{Po}^+$ cations reveal clearly that under 
the canonical framework, the spatial orientation of the individual Cartesian magnetisation components are not unconstrained. For instance, in the case of $\text{H}_2\text{O}^+$, where SOC is essentially absent, the specific projection alignment of the components changes arbitrarily upon rotation. 
This represents a fundamental difference from the collinear case, where the magnetisation vector is strictly forced to lie along the pre-defined longitudinal $z$-axis.
Crucially, because the canonical approach evaluates the exchange-correlation functional (and its corresponding potential) solely using the scalar magnitude of the magnetisation vector field ($m$), this is the only magnetic variable that explicitly determines the total electronic energy. As a consequence, $m$ is strictly invariant before and after spatial rotation, whereas the individual Cartesian components ($m_x, m_y, m_z$) retain a degree of arbitrariness. While these Cartesian components can vary significantly across successive self-consistent field (SCF) iterations, the scalar norm $|\mathbf{m}|$ remains well-behaved and tightly converged throughout the SCF optimisation process.

\begin{table}[h!]
    \centering
    \begin{tabular}{ccccc}
    \toprule
    &  {\mathtxt{H_2O^+} (0°)} & {\mathtxt{H_2O^+} (90°)} &  {\mathtxt{H_2Po^+} (0°)} & {\mathtxt{H_2Po^+} (90°)} \\
    \midrule
    \mathtxt{m_x} &0.022 & 1.089 & 0.120 & 0.129   \\
   \mathtxt{m_y} & 0.036 & 0.036 &0.243 & 0.119 \\
   \mathtxt{m_z} & 1.089 & 0.032 &1.008 & 1.024  \\
   \mathtxt{|m|} & 1.090 & 1.090 & 1.114 & 1.113 \\
   \bottomrule
    \end{tabular}
    \caption{Values of integrated magnetisation vector components and absolute value for \mathtxt{H_2O^+} and \mathtxt{H_2Po^+}, calculated at the relativistic level with BERTHA (canonical LDA). Both 0° and 90° geometries are reported.}
    \label{tab:mag_canonical_lda}
\end{table}

%\begin{figure}
%    \centering
%    \includegraphics{H2Po_0_rel_can.png}
%    \includegraphics{H2Po_90_rel_can.png}    
%    \caption{Isosurfaces of the magnetisation vector components for \mathtxt{H_2Po^+} at 0° and 90° geometry, calculated with canonical LDA in BERTHA. Components \mathtxt{m_x}, \mathtxt{m_y} and \mathtxt{m_z} in red, green and blue, respectively.}
%    \label{fig:mag_h2po_can_lda}
%\end{figure}

\subsection{Dissociation of H2 molecule}
As stated in Section 2, the non-relativistic limit of the 
four-component DKS formalism both in collinear and non-collinear frameworks yields the standard unrestricted Kohn-Sham (UKS).
The implementation of these collinear and non-collinear approaches within the BERTHA code is expected to enable the description of bond dissociation mechanisms along the potential energy surface. The previously available density-only method could not describe spin-polarised systems, effectively operating as a restricted closed-shell formalism that is fundamentally unsuited to open-shell systems. 
In contrast, the newly implemented collinear and canonical non-collinear frameworks are 
expected to follow the behaviour of unrestricted methods, as the explicit introduction of the magnetisation density vector enables the unconstrained representation of unpaired spins. 
The homolytic dissociation of the hydrogen molecule ($\text{H}_2$) serves here as a prototypical benchmarking. 

\begin{figure}
    \centering
    \includegraphics{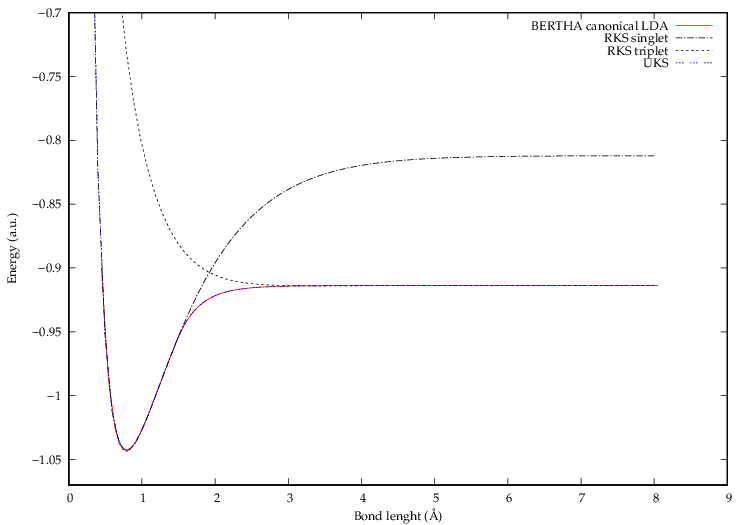}
    \includegraphics{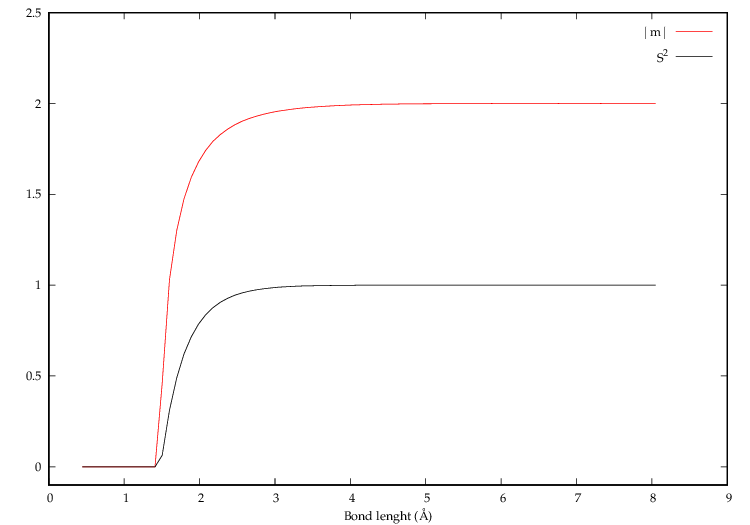}
    \caption{Dissociation curve of \mathtxt{H_2}. Restricted Kohn-Sham (RKS) singlet/triplet and Unrestricted Kohn-Sham (UKS) calculated with ORCA using broken-symmetry approach, canonical non-collinear LDA calculated with BERTHA. Functional used: LDA. a) dissociation plot, b) magnetisation and eigenvalue of \mathtxt{S^2} (canonical and UKS respectively).}
    \label{fig:diss_mag}
\end{figure}

In Figure \ref{fig:diss_mag}a, the non-relativistic Kohn-Sham dissociation curves computed with ORCA package  are compared against that obtained via the canonical non-collinear LDA implementation in BERTHA. The singlet and triplet restricted Kohn-Sham (RKS) profiles correctly describe the short-range and long-range asymptotic behaviours, respectively. However they fail at the medium distances (1.5-3.0 \AA). Meanwhile, the UKS curve smoothly interpolates across the entire distance range.
The canonical non-collinear LDA curve calculated at DKS level is completely superimposed onto the UKS reference profile, confirming that evaluating the functional using the scalar magnetisation vector provides a reliable description of open-shell configurations that is directly comparable to non-relativistic  unrestricted schemes.
This strict relationship is further substantiated by Figure \ref{fig:diss_mag}b, which monitors the scalar magnetisation ($m$) alongside the expectation value of the total spin operator ($\langle S^2 \rangle$). The $\langle S^2 \rangle$ values reflect the degree of spin contamination inherent to the unrestricted method; since a broken symmetry solution 
is used in ORCA to converge at large distance, any deviation from zero indicates a mixing of different electronic spin configurations. The monotonic growth of $\langle S^2 \rangle$ as a function of the internuclear distance is in full agreement with the emergence of a multi-reference broken-symmetry state during homolytic cleavage. This behaviour is mirrored by the total integrated magnetisation module, $m$, which evolves smoothly from 0 to 2. This transition marks the shift from a spin-paired singlet at equilibrium to a fully dissociated state with two independent, unpaired electrons localised on separate nuclei, which is energetically identical to the broken symmetry solution in ORCA but magnetically quite different. It is interesting to note no constraints are imposed on the spin state during bond stretching and the magnetisation is allowed to evolve freely. Nevertheless, the current non-collinear framework inherently suffers from spin contamination, closely mirroring the behaviour of standard non-relativistic unrestricted methodologies. A potential remedy for this issue involves adapting restricted open-shell strategies, such as the Restricted Open-shell Kohn-Sham (ROKS) ansatz \cite{Frank1998,Filatov1998}. In a relativistic four-component framework, where spin-orbit coupling mixes spatial and spin degrees of freedom, such an adaptation would yield the Kramers-restricted open-shell averages\cite{Wang2003_Polarization}.

\section{Conclusions and Perspectives}

In this work, we present the collinear and non-collinear (canonical and Scalmani-Frisch) expressions for evaluating the matrix elements within the relativistic four-component Dirac-Kohn-Sham (DKS) theory, based on G-spinors as implemented in the BERTHA code, and using standard non-relativistic LDA and GGA exchange-correlation functionals. The work provides a unified formalism, which paves the way for a systematic comparison of different methods using a single code to properly control all numerical details.
Furthermore, we have expanded the capabilities of the BERTHA code beyond its original closed-shell, density-only approximation to include both collinear and non-collinear frameworks, thereby enabling the rigorous simulation of spin-polarized and open-shell systems. 

The collinear (LDA and GGA) and canonical non-collinear (LDA functional) methods have been implemented in the BERTHA code, and illustrative calculations have been carried out. We selected the chalcogen hydride series, $\mathrm{H_2X}$ ($\mathrm{X = O, S, Se, Te, Po}$), which exhibits a systematically increasing spin-orbit coupling (SOC) effect, evaluating their ionisation energies in different spatial orientations.
The analysis shows that both the collinear and non-collinear approaches correctly reduce to the non-relativistic Kohn-Sham theory in the absence of significant spin-orbit coupling. However, in the presence of SOC, significant breaking of rotational invariance is observed in the collinear scheme for molecules containing heavier atoms, with an increasing effect on the ionisation energy when descending along the group (from $10^{-6}$ to $10^{-3}$ eV). By visualising the magnetisation vector maps and comparing the $m_z$ isosurfaces to the total magnitude $m$ isosurfaces for \mathtxt{H_2Po}, we can easily explain the breaking of rotational invariance in the collinear approach, which, by discarding the $m_x$ and $m_y$ components, is reliable only when the system has no transverse components (i.e. when $m_x$ and $m_y$ can be neglected so that they do not contribute to the energy of the system).
The analysis of the non-collinear canonical approach tested on \mathtxt{H_2O} and \mathtxt{H_2Po} shows a restoration of rotational invariance. The only factor influencing the final energy of the system is the absolute value of the magnetisation $m$, while the three components of the vector can vary independently without significant effects.

In addition, the dissociation of \mathtxt{H_2} was analysed with the canonical (LDA) approach, demonstrating that it can correctly describe the change in spin state from a closed-shell singlet to an open-shell triplet along the potential energy surface, which is accurately reproduced, with the magnetisation vector value changing from 0 (no unpaired electrons) to 2 (two unpaired electrons of the same spin, one on each hydrogen atom), consistent with the change in spin state. Although this approach is expected to suffer of spin contamination, similar to non-relativistic unrestricted KS methods, in our implementation no magnetisation  constraints are imposed  on the system. 

The advancements presented in this work lay the foundations for several future developments addressing currently unresolved aspects of the formalism.
They will focus on the implementation of canonical non-collinear and Scalmani-Frisch scheme for  GGAs within the BERTHA code. Since GGA functionals are notoriously susceptible to numerical instabilities, resolving these challenges is a prerequisite for a robust implementation: this effort is currently underway in our group.
Beyond these developments, it would be highly desirable to extend the density fitting procedure to these new frameworks. Since these methods currently rely on numerical integrals involving G-spinor amplitudes and their gradients, incorporating density fitting will significantly reduce the computational cost. This scheme will be integrated within the real-time Dirac-Kohn-Sham (DKS) implementation \cite{DeSantis2020} already available in BERTHA: such integration will facilitate the investigation of the time evolution of magnetisation and the analysis of spin dynamics. Ultimately, it will provide an ideal framework to evaluate how different approximations impact the actual spin dynamics system, particularly regarding their ability to satisfy exact theoretical conditions such as local and global torque properties.

\appendix
\section{Equivalence of variables of the theory and polarized variables}\label{equivalence-framework}

Independent variables of relativistic spin-density functional theory($\rho,\mathbf{m}$) and spin-polarized variables ($\rho_+,\rho_-,\gamma_{++},\gamma_{--},\gamma_{+-}$) are equivalent. The choice between the two just depends on the most comfortable framework to work in and the type of functional libraries used. The change from one to the other is possible by using the chain rule:

\begin{equation}
    K_{\mu\nu}^{TT}= \frac{\delta E_{xc}}{\delta D_{\mu\nu}^{TT}} 
    = \int \sum_{i,j} \frac{\delta \epsilon_{xc}}{\delta u_i}\frac{\delta u_i}{\delta V_j}\frac{\delta V_j}{\delta D_{\mu\nu}^{TT}}
\end{equation}

It is briefly demonstrated below for the collinear LDA expressions. Starting by declaring the equivalence

\begin{equation}
    \epsilon_{xc}[\rho, m_z] \longleftrightarrow \epsilon'_{xc}[\rho_+, \rho_-] 
\end{equation}

one then applies the chain rule and substitutes appropriately:

\begin{eqnarray}\nonumber
&& \frac{\partial \epsilon_{xc}[\rho, m_z]}{\partial \rho} = \frac{\partial \epsilon'_{xc}[\rho_+, \rho_-]}{\partial \rho_+} \frac{\partial \rho_+}{\partial \rho} + \frac{\partial \epsilon'_{xc}[\rho_+, \rho_-]}{\partial \rho_-} \frac{\partial \rho_-}{\partial \rho} = \\ 
&& = \frac{1}{2}\frac{\partial \epsilon'_{xc}[\rho_+, \rho_-]}{\partial \rho_+} + \frac{1}{2}\frac{\partial \epsilon'_{xc}[\rho_+, \rho_-]}{\partial \rho_-}
\end{eqnarray}

\begin{eqnarray}\nonumber
&& \frac{\partial \epsilon_{xc}[\rho, m_z]}{\partial m_z} = \frac{\partial \epsilon'_{xc}[\rho_+, \rho_-]}{\partial \rho_+} \frac{\partial \rho_+}{\partial m_z} + \frac{\partial \epsilon'_{xc}[\rho_+, \rho_-]}{\partial \rho_-} \frac{\partial \rho_-}{\partial m_z} = \\
&&= \frac{1}{2}\frac{\partial \epsilon'_{xc}[\rho_+, \rho_-]}{\partial \rho_+} - \frac{1}{2}\frac{\partial \epsilon'_{xc}[\rho_+, \rho_-]}{\partial \rho_-}
\end{eqnarray}

By inserting the results of this step in the equation derived for the LDA as a function of $\rho$ and $m_z$ 

\begin{equation}
 K_{\mu\nu}^{TT} = \frac{\delta E_{xc}}{\delta \matel} \nonumber = \funcres{\rho} \frac{\partial \rho}{\partial \matel}\dr + \funcres{m_z} \frac{\partial m_z}{\partial \matel}\dr
\end{equation}

one obtains the LDA as a function of $\rho_+$, $\rho_-$:

\begin{eqnarray}
     K_{\mu\nu}^{TT} && = \int \frac{\partial \epsilon_{xc}[\rho,m_z]}{\partial \rho} \ [\chi_{\mu}^{\dag T}\chi_{\nu}^{T}]\dr + \\\nonumber
     && + \int \frac{\partial \epsilon_{xc}[\rho,m_z]}{\partial m_z}  [\chi_{\mu}^{\dag T} \sigma_z \chi_{\nu}^{T}]\dr = \\\nonumber
    && = \int [\frac{1}{2}(\frac{\partial \epsilon'_{xc}[\rho_+, \rho_-]}{\partial \rho_+} + \frac{\partial \epsilon'_{xc}[\rho_+, \rho_-]}{\partial \rho_-})] [\chi_{\mu}^{\dag T} \chi_{\nu}^{T}]\dr + \\\nonumber
    && + \int [\frac{1}{2}(\frac{\partial \epsilon'_{xc}[\rho_+, \rho_-]}{\partial \rho_+} - \frac{\partial \epsilon'_{xc}[\rho_+, \rho_-]}{\partial \rho_-})] [\chi_{\mu}^{\dag T} \sigma_z \chi_{\nu}^{T}] \dr = \\\nonumber
    && = \int \frac{\partial \epsilon'_{xc}[\rho_+,\rho_-]}{\partial \rho_+} \frac{1}{2}[\chi_{\mu}^{\dag T} \chi_{\nu}^{T} + \chi_{\mu}^{\dag T} \sigma_z \chi_{\nu}^{T}]\dr + \\\nonumber
    && \int \frac{\partial \epsilon'_{xc}[\rho_+,\rho_-]}{\partial \rho_-} \frac{1}{2}[\chi_{\mu}^{\dag T} \chi_{\nu}^{T} - \chi_{\mu}^{\dag T} \sigma_z \chi_{\nu}^{T}]\dr
\end{eqnarray}

This proves that the two approaches are completely equivalent, and the choice of framework is simply to better complement the exchange-correlation libraries chosen.

\section{Proof of local and global torque theorems}\label{proof_torque}

The global and local torque theorems are exact properties of the exchange-correlation functionals \cite{capelle-vignale}, that have to be satisfied by a prospective non-collinear approach to properly describe spin dynamics. 
The following proof is detailed in the literature \cite{Petrone2018} for the Scalmani-Frisch approach, which satisfies both theorems, and is extended here also for the canonical, which instead does not correctly give a non-vanishing local torque.
The definition of local torque is:

\begin{equation}
    \mathscr{T}_i(\mathbf{r}) = \sum_{jc}\epsilon_{ijc} m_j B^c_{xc}
\end{equation}

where $\epsilon_{ijc}$ is a 3-rank Levi-Civita tensor, and it is an equivalent way to write the vector product. The tensor has the following properties \cite{tyldesley1975introduction}:

\begin{equation}
    \epsilon_{ijk} = 
    \left\{
    \begin{array}{rl}
         +1 \  & \text{if} \ (i,j,k) \ \text{is} \ (1,2,3), \ (2,3,1), \ \text{or} \ (3,1,2) \\
         -1 \  & \text{if} \ (i,j,k) \ \text{is} \ (3,2,1), \ (1,3,2), \ \text{or} \ (2,1,3) \\
         0 \  & \text{if} \ i=j, \ j=k, \ k=i
    \end{array}
        \right.
\end{equation}

where $1,2,3=x,y,z$. In the following work $i$ is always 1 as it does not vary within the expression. The definition of global torque follows as:

\begin{equation}
    \mathscr{T}_{i \ global} = \int \mathscr{T}_i(\mathbf{r}) \ d^3 r .
\end{equation}

The Euler-Lagrange theorem can be applied to $B^c_{xc}$:

\begin{equation}
    \mathcal{B}_{xc}^c = \frac{\delta E_{xc}}{\delta m_c} = \frac{\delta E_{xc}}{\delta m_c} - \nabla \cdot \frac{\delta E_{xc}}{\delta \nabla m_c} = \mathcal{B}_{xc}^{m_c} + \mathcal{B}_{xc}^{\nabla}.
\end{equation}

So the torque can also be split into two contributions, due to the fact that it depends linearly on $B^c_{xc}$:

\begin{equation}
    \mathscr{T}_i(\mathbf{r}) = \mathscr{T}_i(\mathbf{r})^{m} + \mathscr{T}_i(\mathbf{r})^{\nabla}
\end{equation}

\begin{eqnarray}
    \mathscr{T}_i(\mathbf{r})^{m} = \sum_{jc}\epsilon_{ijc} m_j B^{m_c}_{xc} \quad \quad \mathscr{T}_{i \ global}^{m} = \int \mathscr{T}_i(r)^{m_c} \ d^3 r \\\nonumber
    \mathscr{T}_i(\mathbf{r})^{\nabla} = \sum_{jc}\epsilon_{ijc} m_j B^{\nabla_c}_{xc} \quad \quad \mathscr{T}_{i \ global}^{\nabla} = \int \mathscr{T}_i(r)^{\nabla_c} \ d^3 r
\end{eqnarray}

For the LDA case, the definition of $B^c_{xc}$ (which can be derived by developing the functional derivative with respect to the components of the magnetization vector):

\begin{equation}
    \mathcal{B}_{xc}^{m_c} = \frac{1}{2}\frac{m_c}{m}[\frac{\delta \epsilon_{xc}}{\delta \rho_+} - \frac{\delta \epsilon_{xc}}{\delta \rho_-}]
\end{equation}

can be substituted in the torque:

\begin{equation}
    \mathscr{T}_i(\mathbf{r})^{m} = \sum_{jc}\epsilon_{ijc} m_j \{ \frac{1}{2}\frac{m_c}{m}[\frac{\delta \epsilon_{xc}}{\delta \rho_+} - \frac{\delta \epsilon_{xc}}{\delta \rho_-}] \} .
\end{equation}

Looking at $\sum_{jc}\epsilon_{ijc} m_jm_c$, one can observe that the term is always null. Due to the nature of the Levi-Civita tensor, only two cases have to be considered, as all other would be zero due to $i$ being fixed:

\begin{eqnarray}   
    \epsilon_{123}m_2m_3 = + \ m_2m_3 \\\nonumber
    \epsilon_{132}m_3m_2 = - \ m_3m_2
\end{eqnarray}

Meaning that their sum would equal zero:

\begin{equation}
    \sum_{jc=1,2,3}\epsilon_{ijc} m_jm_c \ = \ 0
\end{equation}

This shows that $\mathscr{T}_i(\mathbf{r})^{m}$ is always zero, consequently the global torque is also zero:

\begin{equation}
    \mathscr{T}_{i \ global}^{m} = \int \mathscr{T}_i(\mathbf{r})^{m} \ d^3 r \ = \ 0
\end{equation}

demonstrating that, when the exchange-correlation functional depends only on the density and the magnetisation, as in the LDA case, the local torque is null.
This is not necessarily true for the torque depending on the gradient, and it will be demonstrated for the canonical and Scalmani definition of the GGA variables.

\subsection{Canonical approach}
Verifying the value of local and global torque for GGA functionals requires evaluating an additional term related to the gradient, $B^{\nabla_c}_{xc}$:

\begin{eqnarray}
    \mathcal{B}_{xc}^{\nabla_c} = -\frac{1}{2}\frac{m_c}{m}\nabla \cdot \{[\frac{\delta \epsilon_{xc}}{\delta \gamma_{++}} 2\nabla\rho_+ - \frac{\delta \epsilon_{xc}}{\delta \gamma_{--}} 2\nabla\rho_- - \frac{\delta \epsilon_{xc}}{\delta \gamma_{+-}} (\nabla\rho_+ - \nabla\rho_-)] \}.
\end{eqnarray}

Substituting the term in the expression for the local torque, one gets:

\begin{eqnarray}
    && \mathscr{T}_i(\mathbf{r})^{\nabla} = \sum_{jc}\epsilon_{ijc} m_j(r) \{ -\frac{m_c}{2m} \nabla \cdot [\frac{\delta \epsilon_{xc}}{\delta \gamma_{++}} 2\nabla\rho_+ - \\ \nonumber
    && - \frac{\delta \epsilon_{xc}}{\delta \gamma_{--}} 2\nabla\rho_- - \frac{\delta \epsilon_{xc}}{\delta \gamma_{+-}} (\nabla\rho_+ - \nabla\rho_-)]\}
\end{eqnarray}

It can be seen that the term to evaluate is, as before, $\sum_{jc}\epsilon_{ijc} m_jm_c$, which still amounts to zero. As such the local torque is zero also for the GGA part in the canonical approach. The global torque is consequently also null, as it should. However, the lack of local torque amounts to a lacklustre description of the behaviour of the system: a functional that does not satisfy the non-vanishing local torque theorem is not sensitive to changes in the direction of the magnetisation and does not properly describe the spin dynamics of the system.

\subsection{Scalmani-Frisch approach}

The Scalmani-Frisch approach has the important characteristic, which makes it an improvement over the canonical approach, of correctly satisfying the non-vanishing local torque theorem, due to its unique redefinition of the variables. Proof of these theorems will be presented below. \cite{Petrone2018}
Starting with the definition of local torque for the GGA terms

\begin{equation}
    \mathscr{T}_i(\mathbf{r})^{\nabla} = \sum_{jc}\epsilon_{ijc} m_j B^{\nabla_c}_{xc}
\end{equation}

Analogously to the canonical method, one needs to substitute the expression of the $B^{\nabla_c}_{xc}$ potential for the Scalmani-Frisch approach:

\begin{eqnarray}\label{eq:b-pot-rearranged}
    B^{\nabla_c}_{xc} && = -\nabla \cdot \frac{1}{2} \nabla m_c \{ \frac{\delta \epsilon_{xc}}{\delta \gamma_{++}}[1 + \\ \nonumber
    && +\frac{\partial}{\partial \nabla m_c}\frac{f_{\nabla}}{2}\sqrt{(\nabla\rho \cdot \nabla\mathbf{m}) \circ (\nabla\rho \cdot \nabla\mathbf{m})} + \\ \nonumber
    && + \frac{f_{\nabla}}{2\bigstar(\nabla\rho \cdot\nabla m_c)\nabla\rho}] + \\ \nonumber
    && \frac{\delta \epsilon_{xc}}{\delta \gamma_{--}}[1 - \frac{\partial}{\partial \nabla m_c}\frac{f_{\nabla}}{2}\sqrt{(\nabla\rho \cdot \nabla\mathbf{m}) \circ (\nabla\rho \cdot \nabla\mathbf{m})} + \\ \nonumber
    && + \frac{f_{\nabla}}{2\bigstar}(\nabla\rho \cdot\nabla m_c)\nabla\rho] + \frac{\delta \epsilon_{xc}}{\delta \gamma_{+-}} \}
\end{eqnarray}

with $\bigstar$ (star) as a shorthand for  $\sqrt{(\nabla\rho \cdot \nabla\mathbf{m}) \circ (\nabla\mathbf{m} \cdot \nabla\rho)}$.
It is clear that, when substituted in the previous expression, there is not a term "capable" of contributing to the Levi-Civita tensor to nullify the expression: the local torque for the Scalmani-Frisch formulation is consequently not zero: 

\begin{equation}
    \mathscr{T}_i(\mathbf{r})^{\nabla} \neq 0
\end{equation}

Now, it must be proven that the global torque is still zero, whose definition is still:

\begin{equation}
    \mathscr{T}_{i \ global}^{\nabla} = \int \mathscr{T}_i(r)^{\nabla_c} \ d^3 r = \int \sum_{jc}\epsilon_{ijc} m_j B^{\nabla_c}_{xc} \ d^3 r
\end{equation}

This is done by applying Green's theorem to the potential of Equation \ref{eq:b-pot-rearranged} substituted in the global torque expression

\begin{eqnarray}\nonumber
    && \int \sum_{jc}\epsilon_{ijc} m_j B^{\nabla_c}_{xc} \ d^3 r = - \int \sum_{jc}\epsilon_{ijc} \nabla m_j \frac{1}{2} \nabla m_c \{ \frac{\delta \epsilon_{xc}}{\delta \gamma_{++}}[1 \\
    && + \frac{\partial}{\partial \nabla m_c}\frac{f_{\nabla}}{2}\sqrt{(\nabla\rho \cdot \nabla\mathbf{m}) \circ (\nabla\rho \cdot \nabla\mathbf{m})} + \\ \nonumber
    && + \frac{f_{\nabla}}{2\bigstar}(\nabla\rho \cdot\nabla m_c)\nabla\rho] + \frac{\delta \epsilon_{xc}}{\delta \gamma_{--}}[1 - \\ \nonumber
    && - \frac{\partial}{\partial \nabla m_c}\frac{f_{\nabla}}{2}\sqrt{(\nabla\rho \cdot \nabla\mathbf{m}) \circ (\nabla\rho \cdot \nabla\mathbf{m})} + \\ \nonumber
    && + \frac{f_{\nabla}}{2\bigstar}(\nabla\rho \cdot\nabla m_c)\nabla\rho] + \frac{\delta \epsilon_{xc}}{\delta \gamma_{+-}} \} \ d^3 r
\end{eqnarray}

In this case, it is clear that the condition to nullify the expression is satisfied, as one has:

\begin{equation}
    \sum_{jc}\epsilon_{ijc}\nabla m_j \nabla m_c
\end{equation}

which means that only two opposite cases can exist, which make it so the sum equals zero

\begin{equation}
    \begin{array}{c}
    \epsilon_{123}\nabla m_2 \nabla m_3 = + \nabla m_2 \nabla m_3 \\
    \epsilon_{123}\nabla m_3 \nabla m_2 = - \nabla m_3 \nabla m_2
    \end{array} \quad \Longrightarrow \quad \sum_{jc}\epsilon_{ijc}\nabla m_j \nabla m_c=0
\end{equation}

and the zero-global torque theorem is satisfied even though the local torque is non-null.

\bmhead{Supplementary information}
Optimized geometries, report of the numerical test of collinear and density only approximations for a closed shell molecule, collinear and non-relativistic UKS comparison and  integrated magnetization components for H$_2$O$^+$ and H$_2$Po$^+$. 
%Authors reporting data from electrophoretic gels and blots should supply the full unprocessed scans for key as part of their Supplementary information. This may be requested by the editorial team/s if it is missing.
%
%Please refer to Journal-level guidance for any specific requirements.

\end{document}